\documentclass[11pt,a4paper]{article}

\usepackage[margin=1in]{geometry}
\usepackage{amsmath,amssymb,mathtools,bm}
\usepackage{microtype}
\usepackage{booktabs}
\usepackage{xcolor}
\usepackage{array}
\usepackage{enumitem}
\usepackage[numbers,sort&compress]{natbib}
\usepackage[colorlinks=true,linkcolor=blue,citecolor=blue,urlcolor=blue]{hyperref}

\newcommand{\AdS}{\mathrm{AdS}}
\newcommand{\cO}{\mathcal O}
\newcommand{\cL}{\mathcal L}
\newcommand{\cN}{\mathcal N}
\newcommand{\cR}{\mathcal R}
\newcommand{\cK}{\mathcal K}
\newcommand{\cC}{\mathcal C}

\newcommand{\RR}{\mathbb R}
\newcommand{\ZZ}{\mathbb Z}
\newcommand{\ellp}{\ell_{\mathrm p}}
\newcommand{\pone}{p_1}
\newcommand{\ptwo}{p_2}

\newcommand{\dd}{\mathrm d}

\newcommand{\vol}{\mathrm{vol}}

\numberwithin{equation}{section}

\title{\bfseries M-theory anomaly localization and holographic $c$-extremization}
\author{Carolina Matt\'e Gregory}
\date{}

\begin{document}

\maketitle

\begin{abstract}
We compute the cubic and $C_3\wedge X_8$ characteristic-class contributions to the trial anomaly
polynomial of supersymmetric $\AdS_3\times M_8$ backgrounds with
$M_8=P(\cO\oplus\cL)\to B_6$.  Including the Abelian currents descending
from $C_3$, we identify the projective invariant $J$ as the denominator
arising from their elimination by $c$-extremization and recover the central charge
$c^{(0)}=-\mathsf{I}_3+3\mathsf{I}_+\mathsf{I}_-/J$.  The $X_8$ push-forward determines the pure
R-symmetry, mixed R--Betti and gravitational anomalies.  Evaluating it at
the classical extremum gives the associated central-charge corrections,
linear in the overall unshifted flux.  The mixed term contributes equally
to both central charges and is required in general.  We evaluate these
formulae for three magnetically flux-quantized backgrounds, including non-polynomial
solutions.  Direct projective fibre integration reproduces the mixed
coefficient obtained by localization.  On the wrapped-M5 $S^4$ bundle,
the same couplings reproduce the independently known field-theory anomaly
through orders $N^3$ and $N$.  Their interpretation as full CFT central charges requires a unitary
SCFT with a normalizable vacuum, suitable supersymmetric boundary conditions,
a complete mixing sector of global currents and control of omitted anomaly
sectors.  A conditional Euclidean relative-equivariant interpretation is
discussed separately.
\end{abstract}

\clearpage
\begingroup
\setcounter{tocdepth}{2}
\footnotesize
\tableofcontents
\endgroup
\clearpage

\section{Introduction}

Supersymmetric $\AdS_3$ backgrounds of M-theory relate two-dimensional
anomalies to eleven-dimensional topological couplings. Alongside the cubic
Chern--Simons term, the three-form potential $C_3$ enters the coupling
\begin{equation}
 C_3\wedge X_8,\qquad X_8=\frac1{192}\left(\pone^2-4\ptwo\right)
 \label{X8-def}
\end{equation}
where $p_1$ and $p_2$ are the first and second Pontryagin classes of the
tangent bundle. Scaling the unshifted magnetic flux by $N$ at fixed topology,
the two terms contribute at orders $N^3$ and $N$, respectively, so anomaly
inflow probes corrections without requiring a higher-derivative solution.

Our starting point is the equivariant inflow framework of Bah, Bonetti,
Minasian and Nardoni \cite{Bah:2019inflow} and the established
$c$-extremization principle \cite{BeniniBobev:2012cext,Benini:2013cda}.
Holographic Betti-vector mixing was previously studied in
\cite{AmaritiToldo:2017}, while geometric $c$-extremization and off-shell
localization for wrapped M5-branes provide related precedents
\cite{CouzensGauntlettMartelliSparks:2019,GauntlettMartelliSparks:2019,BenettiGenoliniGauntlettSparks:2023blocks}.
The cubic-plus-$X_8$ equivariant anomaly form also enters the more recent
relative localization of the M-theory Chern--Simons action
\cite{BenettiGenolini:2026pathintegral}. We use these established ingredients
for the compact projective geometries
\begin{equation}
 M_8=P(\cO\oplus\cL)\longrightarrow B_6,
 \label{projective-geometry}
\end{equation}
where $P$ denotes projectivization, $\cO$ is the trivial bundle and
$\cL$ is a complex line bundle over the six-dimensional base $B_6$.
The $\mathbb{CP}^1$ fibres have a circle action with two fixed sections.
These topologies occur in \cite{Gauntlett:2006ns}. Equivariant localization
gave topological formulae for their leading central charges and gravitational
anomalies \cite{GregoryNotmanSparks:2026localizing}, while
\cite{Gregory:nonpoly} established regular non-polynomial families,
including a certified finite-ratio flux neighbourhood.

For this geometry, the projective intersection ring fixes the trial anomaly
of the $C_3$-derived Betti currents. Their elimination produces a Schur
complement with denominator $J$ and recovers the leading central charge
\begin{equation}
 c^{(0)}=-\mathsf{I}_3+3\frac{\mathsf{I}_+\mathsf{I}_-}{J}.
 \label{intro-leading-c}
\end{equation}
Here $\mathsf{I}_3$, $\mathsf{I}_\pm$ and $J$ are intersection invariants defined
in section~\ref{classical-cs}. The formula was obtained in
\cite{GregoryNotmanSparks:2026localizing}, whereas the trial-current derivation
explains geometrically why $J$ appears.

The principal new subleading result is the projective $C_3\wedge X_8$
trial anomaly, including its mixed R--Betti coefficient. At the classical
extremum this coefficient contributes a common order-$N$ shift to the two
retained central charges, while the gravitational anomaly determines their
difference. Leading-order stationarity makes the subleading displacement of
the trial R-current unnecessary at this order.

Direct fibre integration checks the projective mixed coefficient obtained
from the fixed-section formula. The wrapped-M5 and
Maldacena--Strominger--Witten (MSW) comparisons calibrate known anomaly
terms, while three magnetically flux-quantized projective
backgrounds, including non-polynomial solutions, give explicit applications
of the new correction.

These results concern the retained characteristic-class contributions.
Their interpretation as full CFT central charges also requires a complete
boundary-current sector and control of omitted anomalies, as discussed below.

The paper is organized as follows. Sections~\ref{anomaly-framework}
and~\ref{projective} set out the anomaly conventions and projective flux data.
Section~\ref{classical-cs} derives the leading Schur-complement formula,
section~\ref{X8} computes the $X_8$ correction, and
section~\ref{hd-extremization} performs subleading extremization.
Sections~\ref{M5check} and~\ref{MSW-check} give the wrapped-M5 and MSW
comparisons, followed by the curve-product examples in section~\ref{curves}
and the discussion of scope. Appendices~\ref{projective-technical}--\ref{lorentzian-dictionary}
contain, respectively, the projective reduction, current-sector checks,
wrapped-M5 details, endpoint benchmark and conditional Euclidean formulation.

\section{Anomaly polynomial and conventions}
\label{anomaly-framework}

We consider supersymmetric Lorentzian $\AdS_3\times M_8$ backgrounds of
eleven-dimensional supergravity, allowing both magnetic and electric components
of the four-form field strength. Following the conventions of
\cite[section~2]{GregoryNotmanSparks:2026localizing}, we write the metric
and flux as
\begin{equation}
    \dd s^2_{11,L}=e^{2\lambda}\left[\dd s^2(\AdS_3)+\dd s^2(M_8)\right],
    \qquad G_L=e^{3\lambda}F+
    \vol_{\AdS_3}\wedge e^{3\lambda}f_{\rm el}.
    \label{lorentzian-input}
\end{equation}
Here $\lambda$ is a function on $M_8$, $\dd s^2(\AdS_3)$ is the
unit-radius anti-de Sitter metric, and $\dd s^2(M_8)$ is the unwarped
internal metric. The form $\vol_{\AdS_3}$ is the volume form of the
unit-radius external metric, while $F$ and $f_{\rm el}$ are, respectively,
a four-form and a one-form on $M_8$. Thus, $e^{3\lambda}F$ is the magnetic
part of the physical flux, and the term involving $f_{\rm el}$ describes
its electric component.

The characteristic-class calculation below uses the magnetic cohomology
class rather than the full Lorentzian field strength. Denoting the physical
M-theory four-form by $G_4$, with $G_4=G_L$ on this background, and the
eleven-dimensional Planck length by $\ellp$, we define
$\mathfrak G=G_4/(2\pi\ellp)^3$.
The Bianchi identities give $\dd(e^{3\lambda}F)=0$ and
$\dd(e^{3\lambda}f_{\rm el})=0$, so the first defines the real de Rham class
$\mathfrak g=[e^{3\lambda}F/(2\pi\ellp)^3]\in H^4(M_8;\RR)$, while the
second allows us to introduce a local electric potential. We write
$\widehat{\mathfrak G}$ for an equivariant extension of $\mathfrak g$
that includes the background sources for trial currents. It is not the
full physical four-form $G_4$, although the electric component enters
the supersymmetry lift through the potential discussed in
appendix~\ref{internal-lift-details}. The notation $f_{\rm el}$ also
distinguishes this one-form from the section-flux classes $f_\pm$
introduced in section~\ref{projective}.

The retained characteristic polynomial is
\begin{equation}
 \mathcal I_{12}^{\rm ret}
 =\frac16\widehat{\mathfrak G}^{\,3}
   +\widehat{\mathfrak G}\,\widehat X_8,\qquad
 I_4^{\rm ret}=\left[\int_{M_8}\mathcal I_{12}^{\rm ret}\right]_4.
 \label{compact-anomaly-start}
\end{equation}
Here $\widehat X_8$ is the equivariant extension of the characteristic
polynomial $X_8$ in \eqref{X8-def}, while ``ret'' indicates that only
the specified cubic and $C_3\wedge X_8$ contributions are retained.
We adopt the boundary-anomaly sign convention, opposite to that of brane inflow, and
keep the background gauge and tangent-bundle curvatures formal during integration over
the closed $M_8$. The bracket $[\cdot]_4$ selects the component of external degree four,
which is kept formal during the internal integration. The resulting four-form
encodes the two-dimensional anomalies through descent. At this stage the calculation concerns characteristic classes, while the
physical scope of the retained sector is addressed in section~\ref{hd-extremization}.

Descent is implemented with an auxiliary four-manifold $W_4$ whose boundary is the
three-dimensional spacetime $M_3$, without requiring a supergravity solution on $W_4$.
By contrast, a relative interpretation of the on-shell action would also require global
circle data, extended bilinears and a relative boundary representative, ingredients
considered only in appendix~\ref{relative}.

We use the two-dimensional convention in which $x_A$ are the background
first Chern classes for Abelian global currents labelled by $A$, and $k_{AB}$ are their signed anomaly coefficients. We write $T_2$
for the formal tangent bundle of the two-dimensional theory, whose
Pontryagin class is represented by the corresponding external class on $W_4$.
Then
\begin{equation}
    I_4=\frac12k_{AB}x_Ax_B-\frac{k_{\rm grav}}{24}\pone(T_2),
    \qquad k_{\rm grav}=c_R-c_L,
    \label{2d-anomaly-convention}
\end{equation}
with signed current-algebra coefficients
\begin{equation}
    k_{AB}=+(k_R)_{AB}\ \text{for right-movers},
    \qquad
    k_{AB}=-(k_L)_{AB}\ \text{for left-movers}.
    \label{chirality-level-sign}
\end{equation}
In the projective orientation below the supersymmetric current is labelled
left-moving, hence
\begin{equation}
    c_L=-3k_{RR}.
    \label{cL-kRR}
\end{equation}
Using this relation to interpret the trial anomaly presupposes the boundary-theory
hypotheses collected in section~\ref{trial-space-completeness}, which are not
established by the characteristic-class push-forward alone. For the wrapped-M5
comparison in section~\ref{M5check}, we instead adopt the conventional right-moving
orientation of \cite{Benini:2013cda}. The two conventions are related by reversing the
boundary orientation, which changes the sign of $I_4$ and interchanges left and right.

\section{Projective geometry and equivariant flux data}
\label{projective}

\subsection{The projective bundle}

Consider
\begin{equation}
    E=\cO\oplus\cL,\qquad
    M_8=P(E)\overset{\pi}{\longrightarrow}B_6,
    \qquad \ell=c_1(\cL).
    \label{M8-projective}
\end{equation}
We use the quotient convention for projectivization, so that
$\cO_{P(E)}(1)$ is the tautological quotient line bundle, and set
\begin{equation}
    u_B=c_1\!\left(\cO_{P(E)}(1)\right),
\end{equation}
and the projective-bundle relation is
\begin{equation}
u_B(u_B-\ell)=0.
\label{projective-relation}
\end{equation}

Let the fibre circle act with weight zero on the trivial summand and weight
one on $\cL$.  Its fixed sections, with embeddings $s_\pm:B_6\hookrightarrow M_8$, are
\begin{equation}
    B_-:\ E\twoheadrightarrow\cO,
    \qquad
    B_+:\ E\twoheadrightarrow\cL,
\end{equation}
with normal bundles
\begin{equation}
    N_-=\cL^{-1},
    \qquad
    N_+=\cL.
\end{equation}
Let $\xi$ generate the fibre-circle action and let $w_\xi$ be its formal
equivariant curvature variable. Introducing the associated degree-two
parameter and the equivariant normal-line root,
\begin{equation}
    u_\xi=\frac{w_\xi}{2\pi},
    \qquad
    a_\xi=u_\xi+\ell,
\end{equation}
we obtain the equivariant Euler classes $e_\xi(N_\pm)$,
\begin{equation}
    e_\xi(N_-)=-a_\xi,
    \qquad
    e_\xi(N_+)=a_\xi.
\label{euler-fixed-sections}
\end{equation}
At either fixed section,
\begin{equation}
    TM_8|_{B_\pm}=TB_6\oplus N_\pm,
\end{equation}
so the Whitney product formula gives the equivariant Pontryagin
classes $p_i^\xi$,
\begin{equation}
    \pone^\xi(TM_8)|_{B_\pm}=\pone(TB_6)+a_\xi^2,
\label{p1-fixed}
\end{equation}
\begin{equation}
    \ptwo^\xi(TM_8)|_{B_\pm}
    =\ptwo(TB_6)+\pone(TB_6)a_\xi^2.
\label{p2-fixed}
\end{equation}
Consequently,
\begin{equation}
    X_8^\xi|_{B_\pm}
    =\frac1{192}
    \left[
      \pone(TB_6)^2-4\ptwo(TB_6)
      -2\pone(TB_6)a_\xi^2+a_\xi^4
    \right].
\label{X8-fixed}
\end{equation}
These fixed-section restrictions provide the equivariant curvature data used
in the characteristic-class calculation below.

\subsection{Flux cohomology}

Throughout the projective calculation, we use the \emph{unshifted} magnetic
cohomology class
\begin{equation}
    \mathfrak g
    \equiv
    \left[\frac{G_4}{(2\pi\ellp)^3}\right]
    =f_-+u_B\,\mathfrak n,
    \label{g-projective}
\end{equation}
where
\begin{equation}
    \mathfrak n\in H^2(B_6),
    \qquad
    f_-\in H^4(B_6).
\end{equation}
The class $\mathfrak n$ is the fibre-integrated magnetic flux, while
$f_-$ is its restriction to the negative fixed section. We use $H^k$ for
real de Rham cohomology unless an integral coefficient ring is displayed
explicitly. The physical M-theory quantization law includes a shift
\cite{Witten:1996flux},
\begin{equation}
    \mathfrak q
    =
    \mathfrak g
    -\frac14\pone(TM_8)
    \in H^4(M_8,\ZZ),
    \label{shifted-flux-quantization}
\end{equation}
in our chosen orientation for $G_4$. We shall use the unshifted class $\mathfrak g$ in
the anomaly calculation, while $\mathfrak q$ keeps track of the integral flux lattice.
Let $\pi_*$ denote the Gysin push-forward, namely integration of
cohomology classes over the oriented $\mathbb{CP}^1$ fibre. For the projective bundle,
\begin{equation}
    \pone(TM_8)
    =
    \pi^*\!\left[\pone(TB_6)+\ell^2\right],
\end{equation}
so
\begin{equation}
    \pi_*\pone(TM_8)=0.
\end{equation}
Consequently, the fibre-integrated flux is insensitive to the gravitational
shift,
\begin{equation}
    \pi_*\mathfrak q
    =
    \pi_*\mathfrak g
    =
    \mathfrak n.
    \label{n-shift-invariant}
\end{equation}
At fixed trial sources, the same base-class shift leaves the $X_8$ term
unchanged. A change of charge family may nevertheless require re-expanding
the classical mixing parameters, as explained in section~\ref{flux-grading}.
The gravitational anomaly \eqref{grav-anomaly}, which depends only on
$\mathfrak n$ and topology, is unaffected by this shift.

Restriction to the pole sections yields
\begin{equation}
    \mathfrak g|_{B_-}=f_-,
    \qquad
    \mathfrak g|_{B_+}=f_+,
\end{equation}
with
\begin{equation}
f_+=f_-+\ell\wedge\mathfrak n.
\label{pole-flux-relation}
\end{equation}
In the internal equivariant cohomology of the fibre action, the projective
class restricts as
\begin{equation}
    u_B|_{B_-}=0,
    \qquad
    u_B|_{B_+}=u_\xi+\ell.
\end{equation}
Thus, the same representative gives
\begin{equation}
    \widehat{\mathfrak g}|_{B_+}
    =f_+ +u_\xi\mathfrak n.
\end{equation}
Together, these restrictions specify a convenient reference lift of the magnetic class,
which need not coincide with the supersymmetry polyform \eqref{AdS3-flux-polyform} in a
fixed current basis. In section~\ref{flux-lift-ambiguities}, we explain how a difference
between the two lifts affects the trial anomaly and under which hypotheses its
extremized value is unchanged.

The same projective description gives a useful flux--curvature identity without invoking
the higher-derivative localization argument. The Gysin map obeys
\begin{equation}
    \pi_*(1)=0,
    \qquad
    \pi_*(u_B)=1,
    \qquad
    \pi_*(u_B^m)=\ell^{m-1}\quad (m\geq1).
    \label{gysin-basic}
\end{equation}
For the real rank-two vertical tangent bundle $T_\pi$, one has
\begin{equation}
    \pone(T_\pi)=\ell^2.
\end{equation}
Combining this relation with the Gysin identities gives the exact cohomological
identity
\begin{equation}
    \int_{M_8}\mathfrak g\wedge\pone(TM_8)
    =
    \int_{B_6}\mathfrak n\wedge
    \left[\pone(TB_6)+\ell^2\right].
\label{grav-localization-identity}
\end{equation}
The same result follows by applying fixed-section localization to the internal circle,
where the two contributions proportional to $f_-$ cancel. Thus,
\eqref{grav-localization-identity} is an intrinsic projective-bundle identity,
independent of the higher-derivative localization conjecture.

\section{Leading anomaly and the Schur-complement mechanism}
\label{classical-cs}

\subsection{The Betti Chern--Simons pairing}

Expand the three-form potential along harmonic two-forms $\eta_I$ of
$M_8$, with $I$ indexing the resulting Abelian gauge fields $A^I$ in three
dimensions,
\begin{equation}
    C_3=A^I\wedge\eta_I+\cdots,
\end{equation}
and write a general class as
\begin{equation}
    \eta=\beta+s\,u_B,
    \qquad \beta\in H^2(B_6).
\end{equation}
With $s$ a real coefficient, the two-derivative Chern--Simons term determines the symmetric pairing
\begin{equation}
    \cK(\eta,\eta')
    =\int_{M_8}\mathfrak g\wedge\eta\wedge\eta'.
    \label{K-def}
\end{equation}
The fibre-integrated flux determines the cup-product map
\begin{equation}
    \mathsf{L}_{\mathfrak n}:H^2(B_6)\longrightarrow H^4(B_6),
    \qquad
    \beta\longmapsto\beta\wedge\mathfrak n.
    \label{Lefschetz-map}
\end{equation}
Using \eqref{projective-relation} and \eqref{gysin-basic}, the pairing
in
\begin{equation}
    H^2(M_8)=H^2(B_6)\oplus\RR\,u_B
\end{equation}
has block form
\begin{equation}
    \cK=
    \begin{pmatrix}
      \mathsf{L}_{\mathfrak n} & f_+\\[1mm]
      f_+^{\,T} & \displaystyle\int_{B_6}\ell\wedge f_+
    \end{pmatrix}.
\label{K-block}
\end{equation}
Here and below, a cohomology class in the off-diagonal block denotes the
corresponding Poincar\'e pairing with a base two-form.

Assume first that $\det\mathsf{L}_{\mathfrak n}\ne0$ on the base
sector under consideration, so that the inverse cup-product map exists. Define
\begin{equation}
    a_+=\mathsf{L}_{\mathfrak n}^{-1}(f_+),
\end{equation}
\begin{equation}
    \mathsf{I}_3=\int_{B_6}\mathfrak n^3,
    \qquad
    \mathsf{I}_\pm=\int_{B_6}\mathfrak n\wedge f_\pm,
    \qquad
    J=\int_{B_6}a_+\wedge f_-.
    \label{I-definitions}
\end{equation}
From
\begin{equation}
    f_+=f_-+\mathsf{L}_{\mathfrak n}\ell,
\end{equation}
we obtain
\begin{equation}
\begin{aligned}
    \int_{B_6}\ell\wedge f_+
    -\int_{B_6}\mathsf{L}_{\mathfrak n}^{-1}(f_+)\wedge f_+
    &=-J.
\end{aligned}
\label{Schur-J}
\end{equation}
Therefore
\begin{equation}
    \det\cK=-J\,\det \mathsf{L}_{\mathfrak n}.
\label{detK}
\end{equation}
The further condition $J\ne0$ is logically separate from
$\det\mathsf{L}_{\mathfrak n}\ne0$. Together they are equivalent to
$\det\mathcal K\ne0$ in this splitting and are assumed whenever we
invert the full Betti pairing. Thus, $-J$ is the Schur complement of the
base block, and $J$ acquires a geometric role as the denominator controlling
the elimination of the $C_3$-derived currents in the localization formula.

\subsection{The compact cubic anomaly push-forward}
\label{cubic-relative-fixed-point}

To turn this pairing into the two-derivative trial anomaly, we use compact equivariant
push-forward, with its normalization fixed by the eleven-dimensional Chern--Simons
coupling rather than by a choice of supersymmetric filling.

The normalization
\begin{equation}
    \mathfrak G=\frac{G_4}{(2\pi\ellp)^3},
    \qquad 16\pi G_{11}=(2\pi)^8\ellp^9,
\end{equation}
gives the Euclidean cubic Chern--Simons term on an auxiliary
twelve-dimensional characteristic-class extension $W_{12}$,
\begin{equation}
    I_{E,\mathrm{cubic}}
    =
    2\pi i
    \int_{W_{12}}
    \frac16\mathfrak G^3.
    \label{IE-cubic-normalized}
\end{equation}
The background-source extension of the flux follows the inflow
construction of \cite[section~3]{Bah:2019inflow}.  Let $r=c_1(\cR)$ denote
the trial R-symmetry class and let $x^I$ denote background classes for the
vectors descending from $C_3$.  Introduce
\begin{equation}
    \widehat\ell=\ell+2r,
\end{equation}
and write
\begin{equation}
    \eta_I=\beta_I+s_Iu_B,
    \qquad
    A=f_-+x^I\beta_I,
    \qquad
    B=\mathfrak n+x^Is_I.
    \label{A-B-def}
\end{equation}
We use the same symbol $u_B$ for the equivariant extension of the projective
class, with the normal-line twist encoded by $\widehat\ell$.
We choose the reference equivariant flux class
\begin{equation}
    \widehat{\mathfrak G}=A+u_B B,
    \qquad
    u_B(u_B-\widehat\ell)=0.
    \label{G-AyB}
\end{equation}
Here $\widehat\ell$ is understood as the Chern--Weil image of the
equivariant first Chern class of the projective normal line.  Before applying
the Chern--Weil homomorphism, it contains the nonzero fibre weight of the
canonical R-action, so the fixed-point Euler classes are invertible in the
equivariant ring.

The reference lift and the supersymmetry lift give the same extremized
anomaly provided their degree-two difference is represented in the retained
Betti trial space and the source-free flux constraint holds.  The argument
is made explicit in section~\ref{flux-lift-ambiguities}.  In particular,
setting $x=0$ here chooses a reference current, not necessarily the
superconformal current selected by the Killing spinor.

At the two projective fixed sections, the restrictions are
\begin{equation}
    \widehat{\mathfrak G}_- = A,
    \qquad
    \widehat{\mathfrak G}_+ = A+\widehat\ell B,
    \label{G-fixed-cubic}
\end{equation}
and, with the projective orientation used throughout,
\begin{equation}
    e_\xi(N_-)=-\widehat\ell,
    \qquad
    e_\xi(N_+)=\widehat\ell.
    \label{euler-fixed-cubic}
\end{equation}
The compact fixed-section sum is therefore
\begin{equation}
\begin{aligned}
    \mathcal P
    &\equiv
    \sum_{\pm}
    \frac{1}{6}
    \frac{\widehat{\mathfrak G}_\pm^3}{e_\xi(N_\pm)}
    \\
    &=
    \frac{(A+\widehat\ell B)^3-A^3}{6\widehat\ell}
    \\
    &=
    \frac12A^2B
    +\frac12\widehat\ell AB^2
    +\frac16\widehat\ell^2B^3.
    \label{relative-projective-cubic}
\end{aligned}
\end{equation}
Notice that all inverse Euler classes have cancelled.  Equation
\eqref{relative-projective-cubic} is simultaneously the fixed-section
localization formula and the projective Gysin push-forward
\begin{equation}
    \mathcal P
    =
    \pi_*
    \left(
       \frac16\widehat{\mathfrak G}^3
    \right).
    \label{fixed-equals-gysin}
\end{equation}
This compact Gysin identity provides the cubic input to the
Schur-complement calculation.

Define $\mathcal P_4=[\int_{B_6}\mathcal P]_4$, where the subscript
selects external degree four.  The Chern--Simons descent convention is
\begin{equation}
    I_{E,\mathrm{CS}}
    =
    2\pi i
    \int_{W_4} I_4,
    \qquad
    I_4^{(0)}=\mathcal P_4.
    \label{bulk-boundary-anomaly-dictionary}
\end{equation}
The relative sign of the cubic and $X_8$ terms is fixed in
\eqref{compact-anomaly-start}.  The sign in the central-charge relation
comes from the left-moving convention $c_L=-3k_{RR}$, as stated in
\eqref{cL-kRR}.

Expanding \eqref{relative-projective-cubic} to external degree four gives
\begin{equation}
    \mathcal P_4
    =
    \frac23 \mathsf{I}_3\,r^2
    +h_I\,r x^I
    +\frac12\cK_{IJ}x^Ix^J,
    \label{cubic-pushforward}
\end{equation}
where $h_I$ denotes the R--Betti mixing coefficient along $\eta_I$,
as specified by the linear functional
\begin{equation}
    h(\beta+s\,u_B)
    =
    \int_{B_6}\beta\wedge\mathfrak n^2+2s\mathsf{I}_+.
    \label{h-functional}
\end{equation}
Thus, both the quadratic Betti pairing and the R--Betti mixing functional are
components of the same compact cubic characteristic-class push-forward.

To form a trial R-current, we set
\begin{equation}
    x^I=t^I r.
\end{equation}
Since $I_4^{(0)}=\mathcal P_4$, the coefficient of $r^2$ in the boundary
anomaly polynomial is
\begin{equation}
    \frac12 k_{RR}^{\rm trial}
    =
    \frac23\mathsf{I}_3
    +h^Tt
    +\frac12t^T\cK t.
\end{equation}
Using the left-moving relation \eqref{cL-kRR}, the leading trial central
charge follows directly as
\begin{equation}
    c_{\rm trial}^{(0)}(t)
    =
    -3\left[
       \frac43\mathsf{I}_3+2h^Tt+t^T\cK t
    \right].
    \label{trial-leading}
\end{equation}
On the nonsingular branch with
$\det\mathsf{L}_{\mathfrak n}\ne0$ and $J\ne0$, extremization gives
\begin{equation}
    t_*=-\cK^{-1}h.
    \label{tstar}
\end{equation}
Block inversion of \eqref{K-block} yields
\begin{equation}
    (t_*)_B=\frac{\mathsf{I}_+}{J},
    \label{tstar-y}
\end{equation}
and
\begin{equation}
    h^T\cK^{-1}h
    =
    \mathsf{I}_3-\frac{\mathsf{I}_+^2}{J}.
    \label{hinvKh}
\end{equation}
For the smooth compact source-free backgrounds, the integrated magnetic
flux constraint, derived from the electric equation of motion in
appendix~\ref{flux-lift-ambiguities-details}, is
\begin{equation}
    \mathsf{I}_++\mathsf{I}_-=0.
    \label{primitivity}
\end{equation}
Substituting these identities into \eqref{trial-leading} gives
\begin{equation}
    c^{(0)}
    =
    -\mathsf{I}_3+3\frac{\mathsf{I}_+\mathsf{I}_-}{J}.
    \label{c-leading-final}
\end{equation}
Equation \eqref{c-leading-final} reproduces the localized central charge
in equation (4.46) of \cite{GregoryNotmanSparks:2026localizing}.
Our $\mathfrak n$ and $f_\pm$ correspond to its $n$ and $m_\pm$,
while the twisting line bundle $\cL$ is denoted $L$ there.
Here the compact cubic anomaly supplies a trial-current derivation of the
same expression.  The intersection ring determines the anomaly matrix,
and elimination of the Betti mixing parameters identifies $J$ as its
Schur-complement denominator.  The conditional relative interpretation in
appendix~\ref{relative-cubic-disc} requires a global filling, which is
not used in this algebraic derivation. The compact trial anomaly thus
already identifies the geometric origin of $J$ without this additional
assumption.

\subsection{Reference-current independence}
\label{flux-lift-ambiguities}

The compact anomaly polynomial requires an equivariant lift of the magnetic
flux. Two such lifts may differ by a term proportional to $r$ with an
internal degree-two coefficient, as well as by a purely external four-form.
When that coefficient belongs to the retained Betti trial space, its effect
is a change of reference R-current that can be absorbed by a linear
redefinition of trial sources.  The purely external term drops
out for the smooth compact source-free backgrounds used here because
\begin{equation}
 \int_{M_8}\mathfrak g^2=\mathsf{I}_-+\mathsf{I}_+=0,
 \qquad X_8(TM_8)=0\quad\hbox{in real cohomology}.
 \label{main-lift-ambiguity-vanish}
\end{equation}
The detailed cohomological argument is given in
appendix~\ref{projective-technical}.

The significance of this freedom is transparent in the gauge part of the trial
anomaly,
\begin{equation}
 I_{4,\mathrm{gauge}}
 =\frac12 k\,r^2+r\,b^Tx+\frac12x^T\cK x.
\end{equation}
For nondegenerate $\cK$, changing the reference current alters $(k,b)$
but leaves the extremized coefficient
\begin{equation}
 k-b^T\cK^{-1}b
 \label{main-reference-Schur-invariant}
\end{equation}
invariant. Consequently, it is the extremized anomaly, not its decomposition into
``direct'' and ``mixed'' terms, that has an invariant meaning. If further
current directions are required, the truncated trial space alone does not
establish reference-current independence. Within the stated Betti sector, however,
the source-free condition and the assumption that the difference between lifts lies
in the retained directions make the reference lift sufficient for both the cubic
result and the order-$N$ correction. The Schur complement thereby identifies the
extremized anomaly as the reference-independent quantity.

\section{The \texorpdfstring{$C_3\wedge X_8$}{C3 wedge X8} characteristic-class correction}
\label{X8}

\subsection{Compact push-forward and the two fixed sections}
\label{X8-relative-fixed-point}

The general cubic-plus-$X_8$ equivariant anomaly coupling is established
in the inflow and relative-localization frameworks
\cite{Bah:2019inflow,BenettiGenolini:2026pathintegral}.
Here we evaluate its $X_8$ term on the projective two-section geometry with
the full retained Betti-source dependence. The unit-supercharge normalization gives
\begin{equation}
 \widehat\ell=\ell+2r,\qquad
 p_1(T_\pi)=\widehat\ell^{\,2}
 \label{R-twist}
\end{equation}
(a spin characteristic-class extension realizing this gauging is described
in appendix~\ref{projective-technical}).  Use the flux lift
$\widehat{\mathfrak G}=A+u_B B$ of \eqref{A-B-def}--\eqref{G-AyB}.
The correction is the external degree-four component of
\begin{equation}
 I_4^{X_8}=\left[\int_{M_8}
       \widehat{\mathfrak G}\,\widehat X_8\right]_4,
 \label{exact-anomaly-pushforward}
\end{equation}
and the two factors are equivariantly closed, so ordinary compact localization applies. At
the fixed sections, the normal roots are $\pm\widehat\ell$, while the base and external
tangent bundles agree. Since Pontryagin polynomials depend only on even powers of the
roots, their restrictions coincide and we obtain
\begin{equation}
 s_+^*\widehat X_8=s_-^*\widehat X_8=\widehat X_{8,B}.
 \label{X8-same-at-poles}
\end{equation}
Using \eqref{G-fixed-cubic} and \eqref{euler-fixed-cubic}, the fixed-section
sum gives
\begin{equation}
 \pi_*\bigl(\widehat{\mathfrak G}\widehat X_8\bigr)
 =-\frac{A\widehat X_{8,B}}{\widehat\ell}
   +\frac{(A+\widehat\ell B)\widehat X_{8,B}}{\widehat\ell}
 =B\widehat X_{8,B}.
 \label{X8-fixed-sum-collapse}
\end{equation}
This identity makes the cancellation of the section flux and the base Betti components
in $A$ manifest before extremization. What remains is the equivariant fibre integral
$B=\mathfrak n+\sum_I s_Ix^I$, so that section-flux dependence enters the final answer
only through the classical mixing parameters.

\label{two-section-principle}
The same argument applies to an oriented $S^2$ fibration with exactly two
fixed sections, dual equivariant normal lines with roots $\pm\alpha$,
and an equivariant class $\widehat\Psi$ with equal section restrictions
$\widehat\Psi_B$.  For any equivariant class $\widehat\Gamma$,
\begin{equation}
 \pi_*(\widehat\Gamma\widehat\Psi)
 =\frac{s_+^*\widehat\Gamma-s_-^*\widehat\Gamma}{\alpha}
       \widehat\Psi_B
 =(\pi_*\widehat\Gamma)\widehat\Psi_B.
 \label{two-section-factorization}
\end{equation}
This is the two-section form of the Berline--Vergne--Atiyah--Bott (BVAB)
localization formula
\cite{BerlineVergne:1982,AtiyahBott:1984}.  The internal $X_8$ restriction
is already given by \eqref{X8-fixed}, with $a_\xi=\alpha$.
For additional fixed components, unequal restrictions or orbifold weights, the
localization sum must be adapted accordingly. For the projective geometries
considered here, \eqref{two-section-factorization} reduces the full fixed-section
calculation to a single fibre-integrated class.

\subsection{External-degree expansion and direct fibre evaluation}
\label{projective-direct-check}

Set
\begin{equation}
    b=\pone(TB_6),
    \qquad
    p=\pone(TW_4),
    \qquad
    a=\ell+2r.
\end{equation}
For
\begin{equation}
    TW_{12}^{\rm top}
    \simeq
    TB_6\oplus TW_4\oplus T_\pi
\end{equation}
at the level of stable characteristic classes, the Whitney formulae give
\begin{equation}
    \pone=b+p+a^2,
\end{equation}
\begin{equation}
    \ptwo
    =\ptwo(B_6)+\ptwo(W_4)+bp+(b+p)a^2.
\end{equation}
Expanding \eqref{X8-def}, the pieces needed for a two-dimensional anomaly
polynomial are
\begin{equation}
    [X_8]_{B_6^4\,W_4^4}
    =
    -\frac1{96}(b+\ell^2)p
    +\frac1{24}(3\ell^2-b)r^2,
\label{X8-ext4}
\end{equation}
and
\begin{equation}
    [X_8]_{B_6^6\,W_4^2}
    =
    \frac1{24}\ell(\ell^2-b)r.
\label{X8-ext2}
\end{equation}
These are the internal and external degree components that enter the
compact push-forward.

Including Betti backgrounds $x^I$ associated with
\begin{equation}
    \eta_I=\beta_I+s_Iu_B,
\end{equation}
the $C_3X_8$ sector therefore produces the four-form anomaly polynomial
\begin{equation}
\begin{aligned}
    I_4^{X_8}
    ={}&
    \frac1{24}
    \left[\int_{B_6}\mathfrak n(3\ell^2-b)\right]r^2
    -\frac1{96}
    \left[\int_{B_6}\mathfrak n(b+\ell^2)\right]p
    \\
    &+
    \frac1{24}
    \left[\int_{B_6}\ell(\ell^2-b)\right]
    r\sum_I s_Ix^I.
\end{aligned}
\label{I4-X8-full}
\end{equation}
The coefficients use the normalization of
\eqref{compact-anomaly-start}.

Using \eqref{2d-anomaly-convention}, \eqref{chirality-level-sign} and
\eqref{cL-kRR}, the direct correction to the supersymmetric,
left-moving central charge in the projective orientation is
\begin{equation}
    \delta c_{L,\rm dir}
    =
    -\frac14
    \int_{B_6}\mathfrak n(3\ell^2-b),
\label{delta-cL-direct}
\end{equation}
while the gravitational anomaly is
\begin{equation}
    \delta(c_L-c_R)
    =
    -\frac14
    \int_{B_6}\mathfrak n(b+\ell^2).
\label{grav-anomaly}
\end{equation}
Equation~\eqref{grav-anomaly} agrees with equation (4.49) of
\cite{GregoryNotmanSparks:2026localizing} in the same left-moving
supersymmetry convention. The direct R-symmetry correction
\eqref{delta-cL-direct} is not part of that comparison. The mixed correction drops out of the
left--right difference, which is also insensitive to the projective flux-quantization
shift because $\mathfrak n=\pi_*\mathfrak q$. The corresponding mixed R--Betti anomalies
are
\begin{equation}
    \delta k_{RI}^{X_8}
    =\frac{s_I}{24}\int_{B_6}\ell(\ell^2-b),
    \qquad
    \delta k_{IJ}^{X_8}=0,
\label{R-betti-X8}
\end{equation}
and the absence of a Betti--Betti term follows directly from the linearity of
$C_3\wedge X_8$ in $C_3$.  The three terms in
\eqref{I4-X8-full} therefore have distinct roles. The $r^2$ term gives a
direct R-anomaly, the $rx^I$ term corrects the central charge after
classical mixing is inserted, and the $p_1(TW_4)$ term fixes the
left--right difference independently of mixing.

These coefficients are exact for the specified $C_3\wedge X_8$
characteristic-class term. Their relation to the complete CFT anomaly is
examined in section~\ref{omitted-anomaly-sectors}.

The fixed-section result can also be checked by direct fibre integration
within the same characteristic-class theory. Explicit Chern--Weil
representatives for the projective generator and the vertical tangent class
reproduce \eqref{R-betti-X8} without using the fixed-section sum.
Both methods evaluate the same coupling with the same flux lift and source
normalization.  On the closed test space with
$B_6=(\mathbb{CP}^1)^3$ and unit fibre Betti source, the resulting
characteristic number is
\begin{equation}
 \delta k_{RB}^{X_8}=2.
 \label{projective-closed-mixed-number}
\end{equation}
For a smooth projective threefold base with $\ell=c_1(TB_6)$, the same
coefficient has the index form
\begin{equation}
 \delta k_{R\eta}^{X_8}=2s\,\chi(B_6,\cO_{B_6}).
 \label{R-betti-holomorphic-index}
\end{equation}
Here $\chi(B_6,\cO_{B_6})$ is the holomorphic Euler characteristic of
the base. This gives a topological interpretation of the mixed coefficient,
further developed through the explicit representatives, a closed twelve-dimensional test and a
Riemann--Roch reduction in appendix~\ref{projective-direct-check-details}.
The fixed-section sum, direct fibre integral and closed characteristic-number
test therefore give mutually consistent evaluations of the same mixed
$X_8$ coefficient. These are checks internal to the chosen characteristic
coupling, not separate absolute normalization measurements.

\section{Extremization through order \texorpdfstring{$N$}{N}}
\label{hd-extremization}

We now combine the cubic and $X_8$ coefficients and extremize the resulting
quadratic anomaly polynomial.

\subsection{Block form of the trial anomaly}

To formulate extremization in a way that accommodates additional currents, let $\mathcal
S$ be the space of geometric Abelian trial sources, containing the reference R-symmetry
and any geometric flavour symmetries that are present. We denote by $\mathcal B$ the vector space of Betti-current directions
associated with Abelian gauge fields descending from $C_3$ on the relevant
subspace of $H^2(M_8;\RR)$, and write the degree-two background sources as
\begin{equation}
    u\in\mathcal S\otimes H^2(W_4),
    \qquad
    x\in\mathcal B\otimes H^2(W_4).
\end{equation}
The collective geometric source $u$ is distinct from the projective class
$u_B$ and the equivariant parameter $u_\xi$.

The gauge part of the trial anomaly polynomial has the block form
\begin{equation}
\begin{aligned}
    I_{4,\rm gauge}^{\rm trial}
    ={}&
    \frac12
    \langle u,\mathbb A\,u\rangle
    +
    \langle u,\mathbb B\,x\rangle
    +
    \frac12
    \mathcal K(x,x),
\end{aligned}
\label{trial-anomaly-block-form}
\end{equation}
or equivalently
\begin{equation}
    I_{4,\rm gauge}^{\rm trial}
    =
    \frac12
    \begin{pmatrix}
       u & x
    \end{pmatrix}
    \begin{pmatrix}
       \mathbb A & \mathbb B\\
       \mathbb B^T & \mathcal K
    \end{pmatrix}
    \begin{pmatrix}
       u\\ x
    \end{pmatrix}.
    \label{trial-anomaly-matrix}
\end{equation}
Here $\mathcal K$ is the $C_3$-derived Chern--Simons pairing of
section~\ref{classical-cs}.  The maps $\mathbb A$ and $\mathbb B$ collect,
respectively, the pure geometric and mixed geometric--Betti anomalies.

When the pairing $\mathcal K$ is nondegenerate on the trial Betti sector, the
stationarity equations can be solved explicitly for $x$, giving
\begin{equation}
    x_*=-\mathcal K^{-1}\mathbb B^T u.
    \label{general-xstar}
\end{equation}
Substitution into \eqref{trial-anomaly-block-form} yields the effective
geometric anomaly matrix
\begin{equation}
    \mathbb A_{\rm eff}
    =
    \mathbb A
    -
    \mathbb B\mathcal K^{-1}\mathbb B^T.
    \label{general-Schur}
\end{equation}
This Schur complement is independent of the basis chosen for the Betti
currents.

\subsection{Large-flux expansion and choice of charge family}
\label{flux-grading}

We organize the derivative expansion by scaling the \emph{unshifted}
magnetic flux,
\begin{equation}
 \mathfrak g(N)=N\mathfrak g_0,
 \label{unshifted-flux-ray}
\end{equation}
with topology, normalized flux and current basis fixed.  The source-free
projective backgrounds obey $\int_{M_8}\mathfrak g^2=0$, and this constraint
is preserved by \eqref{unshifted-flux-ray}.  Put
$\Delta=p_1(TM_8)/4$.  The associated shifted charges form the affine family
\begin{equation}
 \mathfrak q(N)=N\mathfrak g_0-\Delta.
 \label{physical-affine-charge-family}
\end{equation}
Flux quantization restricts $N$ to those values for which this class is integral. In the
curve-product geometries considered below, $\Delta$ is itself integral, so an integral
$\mathfrak g_0$ generates integral shifted charges for every positive integer $N$. This
statement concerns the charge lattice and does not replace the existence and regularity
analysis required for the seed background.

A ray of shifted integral charges would instead have
\begin{equation}
 \mathfrak q(N)=N\mathfrak q_0,
 \qquad
 \mathfrak g(N)=N\mathfrak q_0+\Delta.
 \label{shifted-charge-ray}
\end{equation}
Such a ray need not preserve the source-free flux constraint.  Since
$\Delta$ is pulled back from the six-dimensional base, $\Delta^2=0$, and
\begin{equation}
 \int_{M_8}\mathfrak g(N)^2
 =N^2\int_{M_8}\mathfrak q_0^2
  +2N\int_{M_8}\mathfrak q_0\Delta.
 \label{shifted-ray-compatibility}
\end{equation}
For a complete shifted-charge ray to satisfy the source-free condition,
both coefficients in \eqref{shifted-ray-compatibility} must vanish separately.
Such a ray has additional restrictions, including on its retained gravitational
anomaly. Appendix~\ref{charge-rays-details} gives the explicit examples and
explains why derivatives of the leading formula taken transverse to the
constraint locus do not define physical charge-family corrections.
We use the unshifted-flux family \eqref{unshifted-flux-ray} with shifted
quantization imposed through \eqref{physical-affine-charge-family}
throughout the large-flux calculation.

For the two characteristic terms retained in this paper, the scaling is
particularly simple. The cubic term gives an exactly linear Betti block,
while $C_3\wedge X_8$ is linear in the Betti backgrounds and produces
no Betti--Betti correction.  Accordingly,
\begin{equation}
 \mathcal K=N\mathcal K_1,
 \qquad
 \mathbb A=N^3\mathbb A_3+N\mathbb A_1,
 \qquad
 \mathbb B=N^2\mathbb B_2+\mathbb B_0.
 \label{Betti-no-subleading-local}
\end{equation}
The Schur complement therefore has the expansion
\begin{equation}
\begin{aligned}
 \mathbb A_{\rm eff}
 ={}&N^3\left(
 \mathbb A_3-\mathbb B_2\mathcal K_1^{-1}\mathbb B_2^T\right)\\
 &+N\left(
 \mathbb A_1-\mathbb B_0\mathcal K_1^{-1}\mathbb B_2^T
 -\mathbb B_2\mathcal K_1^{-1}\mathbb B_0^T\right)
 +O(N^{-1}).
\end{aligned}
\label{Schur-largeN-retained}
\end{equation}
The Schur-complement expansion shows that higher-derivative effects displace
the normalized mixing parameters only at relative order $N^{-2}$. Since the
leading trial anomaly is stationary, this displacement first changes its
extremized value at order $N^{-1}$. Thus, the entire order-$N$ correction
is obtained by evaluating the $X_8$ anomaly at the \emph{classical}
extremum. For the retained central charges along the unshifted-flux ray,
this gives
\begin{equation}
\begin{aligned}
 c_L^{\rm ret}&=N^3c_L^{(0)}+Nc_L^{(1)}+O(N^{-1}),\\
 c_R^{\rm ret}&=N^3c_L^{(0)}+Nc_R^{(1)}+O(N^{-1}),
\end{aligned}
\label{largeN-structure}
\end{equation}
where $c_L^{(0)}=c^{(0)}(\mathfrak g_0)$. This uses the fixed charge family
described above and a nondegenerate leading Hessian. Appendix~\ref{extremization-details}
extends the argument to a general block expansion, including a possible
$N^{-1}\mathcal K_{-1}$ term. In the retained projective sector, the stationarity
of the leading anomaly is what makes the classical extremum sufficient for the
entire order-$N$ result.

\subsection{Specialization to the projective R--Betti sector}

We now take $\mathcal S$ to be one-dimensional and generated by the reference
R-symmetry background $r$.  Write
\begin{equation}
    \mathfrak n=N\,\mathfrak n_0,
    \qquad
    f_\pm=N\,f_{\pm,0}.
\end{equation}
Then
\begin{equation}
    \mathsf{I}_3=N^3\mathsf{I}_{3,0},
    \qquad
    \mathsf{I}_\pm=N^2\mathsf{I}_{\pm,0},
    \qquad
    J=NJ_0.
    \label{normalized-invariants}
\end{equation}

For $\eta\in\mathcal B$, define its fibre coefficient invariantly by
\begin{equation}
    s(\eta)=\pi_*\eta\in H^0(B_6).
    \label{fibre-functional}
\end{equation}
In the projective splitting $\eta=\beta+s\,u_B$ this is simply the coefficient
$s$.  The leading Betti pairing and R--Betti mixing functional are
\begin{equation}
    \mathcal K=N\mathcal K_1,
    \qquad
    h=N^2h_2,
\end{equation}
with
\begin{equation}
    h_2(\beta+s\,u_B)
    =
    \int_{B_6}\beta\,\mathfrak n_0^2
    +
    2s\mathsf{I}_{+,0}.
    \label{h2-projective}
\end{equation}

Introduce the characteristic numbers
\begin{equation}
\begin{aligned}
    A_0
    &=
    \int_{B_6}
    \mathfrak n_0
    \left(
       3\ell^2-\pone(TB_6)
    \right),
    \\
    T
    &=
    \int_{B_6}
    \ell
    \left(
       \ell^2-\pone(TB_6)
    \right),
    \\
    G_0
    &=
    \int_{B_6}
    \mathfrak n_0
    \left(
       \ell^2+\pone(TB_6)
    \right).
\end{aligned}
\label{A0-T-G0}
\end{equation}
The pure R-symmetry entries of the signed anomaly matrix are
\begin{equation}
    \mathsf a_3=\frac43\mathsf{I}_{3,0},
    \qquad
    \mathsf a_1=\frac1{12}A_0,
    \label{kappa31}
\end{equation}
while the $X_8$ mixed R--Betti functional is
\begin{equation}
    m_0(\eta)
    =
    \frac{s(\eta)}{24}\,T.
    \label{m0-projective}
\end{equation}

It is convenient to rescale the Betti mixing parameter according to
\begin{equation}
    x=N\,\tau\,r,
    \qquad
    \tau\in\mathcal B.
\end{equation}
The signed trial R-anomaly then becomes
\begin{equation}
\begin{aligned}
    k_{RR}^{\rm trial}(\tau)
    ={}&
    N^3
    \left[
       \mathsf a_3
       +
       2h_2(\tau)
       +
       \mathcal K_1(\tau,\tau)
    \right]
    \\
    &+
    N
    \left[
       \mathsf a_1
       +
       2m_0(\tau)
    \right]
    +
    O(N^{-1}).
\end{aligned}
\label{projective-trial-kRR}
\end{equation}

The classical extremum is
\begin{equation}
    \tau_0
    =
    -\mathcal K_1^{-1}h_2.
    \label{tau0}
\end{equation}
In particular, its projective fibre component is
\begin{equation}
    s(\tau_0)
    =
    \frac{\mathsf{I}_{+,0}}{J_0}.
    \label{tau0-fibre}
\end{equation}
The effective signed R-anomaly through order $N$ is therefore
\begin{equation}
\begin{aligned}
    k_{RR}^{\rm eff}
    ={}&
    N^3
    \left[
       \mathsf a_3
       -
       h_2\mathcal K_1^{-1}h_2
    \right]
    \\
    &+
    N
    \left[
       \mathsf a_1
       -
       2m_0
       \left(
          \mathcal K_1^{-1}h_2
       \right)
    \right]
    +
    O(N^{-1}).
\end{aligned}
\label{projective-effective-kRR}
\end{equation}
Equations \eqref{hinvKh} and \eqref{primitivity}, applied to the
normalized flux, give
\begin{equation}
 c_L^{(0)}=-\mathsf{I}_{3,0}+3\frac{\mathsf{I}_{+,0}\mathsf{I}_{-,0}}{J_0}.
 \label{leading-c-from-master-Schur}
\end{equation}
The order-$N$ term in \eqref{projective-effective-kRR} evaluates the
$X_8$ anomaly along that classical direction.

At order $N$, equations \eqref{m0-projective} and
\eqref{tau0-fibre} give
\begin{equation}
    c_L^{(1)}
    =
    -\frac14
    \left[
       A_0
       +
       \frac{\mathsf{I}_{+,0}}{J_0}T
    \right].
    \label{cL1-master}
\end{equation}
The second term represents R--Betti mixing in the reference-current basis \eqref{G-AyB},
although it is only the sum with the direct contribution, evaluated at the extremum,
that is invariant under the reference-current changes discussed in
section~\ref{flux-lift-ambiguities}.

The gravitational anomaly follows from
\begin{equation}
    k_{\rm grav}^{(1)}
    =
    \frac14G_0,
    \qquad
    k_{\rm grav}=c_R-c_L,
    \label{kgrav1-master}
\end{equation}
and therefore
\begin{equation}
    c_R^{(1)}
    =
    -\frac12
    \int_{B_6}
    \mathfrak n_0
    \left(
       \ell^2-\pone(TB_6)
    \right)
    -
    \frac14
    \frac{\mathsf{I}_{+,0}}{J_0}T.
    \label{cR1-master}
\end{equation}

Restoring the overall flux parameter on the unshifted ray, the retained
sector gives
\begin{equation}
\begin{aligned}
    c_L^{\rm ret}
    &=
    N^3c_L^{(0)}
    +
    Nc_L^{(1)}
    +
    O(N^{-1}),
    \\
    c_R^{\rm ret}
    &=
    N^3c_L^{(0)}
    +
    Nc_R^{(1)}
    +
    O(N^{-1}).
\end{aligned}
\label{master-central-charges}
\end{equation}
The remainder $O(N^{-1})$ in these formulae refers to extremization of the retained
local polynomial, not to the combined effect of all possible CFT anomaly sectors. For
the same $N^3$ and $N$ coefficients to describe the full central charges with an
$O(N^0)$ remainder, omitted sectors would have to modify both the extremized R-anomaly
and the gravitational anomaly by at most $O(N^0)$. This cannot be inferred merely from
an order-one coefficient in the fixed Betti basis, since the mixing coefficients
themselves grow with $N$, as discussed in section~\ref{omitted-anomaly-sectors}. The
wrapped-M5 tensor subtraction considered below has an $O(N^0)$ effect because its
geometric mixing parameter remains bounded. Independently of this qualification, the
cubic term has no gravitational anomaly, and the retained sector therefore predicts the
mixing-independent difference
\begin{equation}
    c_R^{\rm ret}-c_L^{\rm ret}
    =
    \frac{N}{4}G_0.
    \label{master-grav-anomaly}
\end{equation}

The results in \eqref{trial-anomaly-matrix}, \eqref{Schur-largeN-retained} and
\eqref{master-central-charges} can thus be viewed as a single signed anomaly matrix
containing both the $N^3$ and $N$ data. Eliminating the $C_3$-derived currents by its
Schur complement is precisely the algebraic operation that implements holographic
$c$-extremization in this sector.

The inverse-matrix formulae assume both that the trial current sector is complete and
that its leading Betti block is invertible, which, when $\mathsf{L}_{\mathfrak n}$ is invertible,
requires $J\ne0$. All three curve-product examples satisfy this condition, while the MSW
specialization in section~\ref{MSW-check} illustrates how a decoupled null direction can
instead be handled directly in the uneliminated polynomial.

\subsection{Power counting for omitted anomaly sectors}
\label{omitted-anomaly-sectors}

The anomaly polynomial retained here captures the contributions of the two specified
eleven-dimensional couplings, but identifying it with the anomaly of the interacting CFT
also requires control over quantum and decoupled sectors in the same current basis, as
in the general inflow framework \cite{Bah:2019inflow}. The issue is particularly
consequential for the projective examples because their Betti mixing coefficients grow
as $O(N)$.

Write the retained gauge polynomial as
$k r^2/2+r b^T x+x^T\mathcal K x/2$, and let
$t=-\mathcal K^{-1}b$ be its extremum.  In the projective sector,
$b=h+m$, where $h=O(N^2)$ and $m=O(1)$, so
$t=N\tau_0+O(N^{-1})$.  Suppose an omitted sector in this same trial
space changes the blocks by $\delta k$, $\delta b$ and
$\delta\mathcal K$.  Differentiating the Schur complement gives
\begin{equation}
 \delta k_{RR}^{\rm eff}
 =\delta k+2t^T\delta b+t^T\delta\mathcal K t
 \qquad\text{to first order in the omitted blocks}.
 \label{omitted-block-first-variation}
\end{equation}
These contractions show how a seemingly small omitted anomaly can be amplified by
mixing. An $O(1)$ mixed coefficient may contribute at $O(N)$, an $O(1)$ Betti--Betti
coefficient at $O(N^2)$, and even an $O(N^{-1})$ Betti--Betti term may affect the
order-$N$ answer, as confirmed by the general expansion in
appendix~\ref{general-Schur-expansion}. The estimates identify potential effects rather
than establishing that any particular omitted sector occurs in these backgrounds.

The displacement of the extremum can be included exactly, namely, setting
$d=\delta b+\delta\mathcal K t$ and assuming
$\mathcal K+\delta\mathcal K$ is invertible, then completing the square gives
\begin{equation}
 \begin{aligned}
 k_{RR,\rm full}^{\rm eff}-k_{RR,\rm ret}^{\rm eff}
 ={}&\delta k+2t^T\delta b+t^T\delta\mathcal K t\\
 &-d^T(\mathcal K+\delta\mathcal K)^{-1}d.
 \end{aligned}
 \label{omitted-block-exact-difference}
\end{equation}
In particular, the final term shows that an $O(1)$ correction to the Betti block can
generate an additional $O(N)$ contribution even beyond first order in the omitted
blocks. A first-order test alone would therefore not control the claimed accuracy.

A sufficient, though not necessary, set of bounds for an $O(N^0)$
central-charge remainder is
\begin{equation}
 \delta k=O(1),\qquad \delta b=O(N^{-1}),\qquad
 \delta\mathcal K=O(N^{-2}),\qquad
 \delta k_{\rm grav}=O(1),
 \label{sufficient-omitted-block-bounds}
\end{equation}
in the fixed basis along the nondegenerate unshifted ray, under which the last term of
\eqref{omitted-block-exact-difference} is $O(N^{-3})$. Larger omitted blocks
could also be harmless if their contractions with the extremal direction
vanished or cancelled, although establishing this would require
further information. Since the necessary control is not available for all possible
sectors, our corrections remain retained-sector predictions rather than full CFT
results through order $N$. The exact identity
\eqref{omitted-block-exact-difference}, together with the gravitational anomaly,
provides a precise diagnostic within a specified trial sector and identifies the
estimates that must be revisited if additional currents enlarge that sector.

\subsection{Trial-current sector and boundary assumptions}
\label{trial-space-completeness}

For a superconformal interpretation we assume that the boundary theory is a unitary
two-dimensional $\mathcal N=(2,0)$ SCFT with a normalizable conformally invariant
vacuum, so that the standard $c$-extremization argument applies \cite[section~2 and
appendix~C]{Benini:2013cda}. We also require a reference R-current assigning unit charge
to the preserved supercharge, flavour currents under which that supercharge is neutral,
and a trial space containing every Abelian infrared global current capable of R-mixing,
without an additional accidental mixing symmetry. Finally, the retained bulk gauge
fields must admit supersymmetric boundary conditions for which the appropriate chiral
boundary components act as nondynamical sources for those global currents.

The boundary-condition hypothesis specifies which boundary theory is being considered,
rather than following automatically from invertibility of the bulk Chern--Simons
pairing. That pairing does not, by itself, construct a supersymmetric boundary completion
or establish vacuum normalizability and completeness of the infrared current sector. In
the Maxwell--Chern--Simons description, the conserved current is associated with the
flat sector, whereas the propagating mode corresponds to a different operator
\cite[section~4]{Benini:2013cda}. Appendix~\ref{trial-current-checks} discusses the
corresponding boundary prescription and signed levels.

Under these assumptions, extremizing the \emph{complete} boundary anomaly
selects the superconformal R-current. Invertibility of the bulk Betti pairing
only makes the retained algebraic extremization well defined, and does not
establish the physical completeness of that current sector. The order-$N$
formulae derived here give the exact contribution of the specified
characteristic-class terms. Identifying them with the full CFT coefficients
also requires control of the omitted sectors described in
section~\ref{omitted-anomaly-sectors}.

For the three product-of-curves examples below, the harmonic-form and
geometric-isometry candidates can be checked explicitly.  Harmonic two-forms
built from $H^1(C_i)\otimes H^1(C_j)$ are orthogonal to the R/factor/fibre
sector and have nondegenerate Chern--Simons blocks for the quoted fluxes.
Their stationary mixing coefficients therefore vanish uniquely.  Continuous
torus translations are flux-obstructed whenever the corresponding magnetic
period is nonzero.  The factor/fibre block is nondegenerate as well because
$N_1N_2N_3\neq0$ and $J\neq0$ in all three examples.

These checks support the chosen bulk trial space for the explicit backgrounds,
although the boundary-theory assumptions remain necessary. The K\"unneth
decomposition, block determinants and torus-flux obstruction are worked out in
appendix~\ref{trial-current-checks}, where they establish nondegeneracy and
decoupling for the additional bulk current sectors examined here.

\section{Wrapped-M5 calibration}
\label{M5check}

The wrapped-M5 $S^4$ bundle provides a calibration against the known
field-theory anomalies \cite{Benini:2013cda,Baggio:2014sna}. Its
normal-bundle and global-angular-form treatment follows the foundational
fivebrane inflow framework \cite{FreedHarveyMinasianMoore:1998,Bah:2019inflow}.
The comparison checks the relative gauge and gravitational coefficients of
the cubic and $C_3\wedge X_8$ terms, together with subleading
extremization, on a topology whose flavour current is geometric rather
than Betti. It does not constitute a second evaluation of the projective
R--Betti coefficient, which was checked in section~\ref{projective-direct-check}.

\subsection{Six-dimensional anomaly and twist}

Let $T$ be the tangent bundle of the six-dimensional M5 worldvolume
and $\cN$ its rank-five real normal bundle. The anomaly polynomial of a
single free tensor multiplet is
\begin{equation}
    I_8[1]
    =
    \frac1{48}
    \left[
       \ptwo(\cN)-\ptwo(T)
       +
       \frac14
       \left(
          \pone(T)-\pone(\cN)
       \right)^2
    \right],
    \label{single-M5-I8}
\end{equation}

For the interacting $A_{N-1}$ theory associated with a stack of $N$
M5-branes, the anomaly is
$(N-1)I_8[1]+(N^3-N)p_2(\cN)/24$.  The Whitney formula for $X_8$
rewrites this as
\begin{equation}
    I_8[A_{N-1}]
    =
    \frac{N^3}{24}\ptwo(\cN)
    +
    N X_8(T\oplus\cN)
    -
    I_8[1].
    \label{M5-X8-rewrite}
\end{equation}

The final term removes the free centre-of-mass tensor, leaving the interacting theory
whose M5-branes wrap a compact K\"ahler surface $M_4$. The symbol $\chi$
will denote its topological Euler characteristic. We use the notation
\begin{equation}
 c=c_1(TM_4),\quad P_1=\int_{M_4}p_1(TM_4),\quad
 \chi=\int_{M_4}c_2(TM_4),\quad C=\int_{M_4}c^2=P_1+2\chi.
 \label{c1sq-M4}
\end{equation}
Let $r$ and $f$ be the first Chern classes of independent R-symmetry and
flavour background line bundles. The parameter $z$ measures the relative
twist of the two complex normal directions, whose Chern roots are
\begin{equation}
\begin{aligned}
    n_A
    &=
    -\frac{1+z}{2}\,c+r+f,
    \\
    n_B
    &=
    -\frac{1-z}{2}\,c+r-f.
\end{aligned}
\label{M5-roots-independent}
\end{equation}

The sum $n_A+n_B=-c+2r$ shows that the internal normal-bundle contribution cancels
$c=c_1(TM_4)$ in the supersymmetric twist, while integrality of the two normal line
bundles restricts $z$. The resulting normal Pontryagin classes are
$p_1(\cN)=n_A^2+n_B^2$ and $p_2(\cN)=n_A^2n_B^2$, and a trial R-current is introduced by
setting $f=\epsilon r$.

\subsection{Bulk reduction and the trial anomaly}
\label{M5-bulk-sphere}

Let $Z_8=M_4\times W_4$, and let $\rho:S(\cN)\to Z_8$ be the
projection of the associated four-sphere bundle, where
$\cN=\RR\oplus(\cL_A)_\RR\oplus(\cL_B)_\RR$ and
$(\cL_{A,B})_\RR$ denote the underlying oriented real two-plane bundles
of complex lines with $c_1(\cL_A)=n_A$ and $c_1(\cL_B)=n_B$.
We use the normalized global angular class $e_4$ with
$\int_{S^4}e_4=2$ and take $\widehat{\mathfrak G}_{S^4}=Ne_4/2$.
The angular-form identities give
\begin{equation}
 \rho_*\!\left(\frac16\widehat{\mathfrak G}_{S^4}^{\,3}\right)
 =\frac{N^3}{24}p_2(\cN),
 \qquad
 \rho_*\!\left(\widehat{\mathfrak G}_{S^4}\widehat X_8\right)
 =N X_8(TZ_8\oplus\cN).
 \label{M5-bulk-eight-form}
\end{equation}

The space $Z_8$ supports the internal and external sources (the derivation and
coefficient table can be found in appendix~\ref{M5-details}).
Reducing the field-theory polynomial gives
$I_4=N^3I_4^{(3)}+NI_4^{(1)}-\int_{M_4}I_8[1]$, with
\begin{equation}
    I_4^{(3)}
    =
    \frac{C}{48}
    \left[
       (3-z^2)r^2
       -4z\,r f
       +(3z^2-1)f^2
    \right],
    \label{M5-I4-N3}
\end{equation}
\begin{equation}
\begin{aligned}
    I_4^{(1)}
    ={}&
    \frac1{48}
    \left[
       (Cz^2-P_1)r^2
       +4Cz\,r f
       +2\chi f^2
    \right]
    \\
    &-
    \frac1{192}
    \left[
       Cz^2+3P_1+2\chi
    \right]
    \pone(TW_4).
\end{aligned}
\label{M5-I4-N1}
\end{equation}

Appendix~\ref{M5-details} derives the same four coefficients directly
from the sphere bundle. The resulting equality with the field-theory
polynomial, before extremization and up to the free tensor subtraction, is
\begin{equation}
 I_{4,\mathrm{bulk}}=N^3 I_4^{(3)}+N I_4^{(1)},\qquad
 I_4-I_{4,\mathrm{bulk}}=-\int_{M_4}I_8[1].
 \label{M5-bulk-field-match}
\end{equation}

In particular,
\begin{equation}
 (k_{FF})^{(1)}=\frac{\chi}{12},\qquad
 k_{\rm grav}^{(1)}=\frac{Cz^2+3P_1+2\chi}{8}.
 \label{FT-FF-N1}
\end{equation}
A flavour--flavour term is allowed here because $f$ couples to a geometric symmetry
within $SO(5)_R$, not to a $C_3$-derived Betti current. Indeed, $H^2(S^4)=0$, so the
sphere bundle has no counterpart of the projective fibre class.

\subsection{Extremization and the finite-\texorpdfstring{$N$}{N} comparison}

In the right-moving convention of \cite{Benini:2013cda},
$c_R^{\rm trial}=3k_{RR}^{\rm trial}$.  Setting $f=\epsilon r$ gives
\begin{equation}
    c_R^{\rm trial}
    =
    N^3\cC_0(\epsilon)
    +
    N\cC_1(\epsilon)
    +
    O(N^0),
    \label{M5-trial-largeN}
\end{equation}
\begin{equation}
    \cC_0(\epsilon)
    =
    \frac{C}{8}
    \left[
       3-z^2
       -4z\epsilon
       +(3z^2-1)\epsilon^2
    \right],
    \label{C0-M5}
\end{equation}
\begin{equation}
    \cC_1(\epsilon)
    =
    \frac18
    \left[
       Cz^2-P_1
       +4Cz\epsilon
       +2\chi\epsilon^2
    \right].
    \label{C1-M5}
\end{equation}

These are the $N^3$ and $N$ coefficients of equation (6.6) of
\cite{Benini:2013cda}, with $d_Gh_G=N^3-N$ and $r_G=N-1$.
For $C(3z^2-1)\ne0$, the leading extremum is
\begin{equation}
 \epsilon_0=\frac{2z}{3z^2-1}.
 \label{epsilon0}
\end{equation}
Stationarity then gives
\begin{equation}
    c_R
    =
    N^3\cC_0(\epsilon_0)
    +
    N\cC_1(\epsilon_0)
    +
    O(N^0),
    \label{M5-extremized}
\end{equation}

The $2\chi\epsilon_0^2$ term in $\cC_1$ includes the geometric
flavour--flavour anomaly.  For $N>1$ and a nonzero denominator, the exact interacting-theory
extremum, including the tensor subtraction, is
\begin{equation}
    \epsilon_*
    =
    \frac{
       2z\,N(N+1)C
    }{
       C(3z^2-1)(N^2+N+1)+2\chi
    }.
    \label{epsilon-exact-M5}
\end{equation}
This is equation (6.7) of
\cite{Benini:2013cda}.  Its expansion is
\begin{equation}
\begin{aligned}
    \epsilon_*
    ={}&
    \frac{2z}{3z^2-1}
    \\
    &-
    \frac1{N^2}
    \frac{2z}{3z^2-1}
    \left[
       1+
       \frac{
          2\chi
       }{
          C(3z^2-1)
       }
    \right]
    +O(N^{-3}).
\end{aligned}
\label{epsilon-largeN-M5}
\end{equation}

The exact expression therefore confirms both the $N^{-2}$ displacement of the mixing
parameter and the rule that the order-$N$ term can be evaluated at the leading extremum.
The particularly simple specialization $z=0$ with $C\ne0$ gives
\begin{equation}
    \delta c_R=-\frac{N}{8}P_1,
    \qquad
    c_R-c_L=\frac{N}{8}\left(3P_1+2\chi\right)+O(N^0).
    \label{4in8-correction}
\end{equation}
This also agrees with the subleading structure of the higher-derivative
wrapped-M5 reductions in \cite{Baggio:2014sna}.

\section{The Calabi--Yau product and the MSW string}
\label{MSW-check}

The MSW string provides an interesting complementary test in which the projective Betti pairing
becomes degenerate. Let $B_6=Y_3$ be a compact Calabi--Yau threefold with holonomy
$SU(3)$, and let $P$ be an integral ample divisor class. In the magnetic-string
construction of \cite{MaldacenaStromingerWitten:1997}, the geometric and flux data are
\begin{equation}
 M_8=S^2\times Y_3,\qquad \ell=0,\qquad
 \mathfrak n=-P,\qquad f_+=f_-=0.
 \label{MSW-flux-data}
\end{equation}
The charge and chirality conventions are those of
\cite[section~4.4]{GregoryNotmanSparks:2026localizing}.  In particular,
$(c_L,c_R)_{\rm here}=(c_R,c_L)_{\rm MSW}$.
One may take a sufficiently large multiple of $P$ to obtain a smooth
very ample representative in the standard wrapped-brane regime.

Since $\mathsf{I}_\pm=J=0$ in this limit, the expressions involving $\mathsf{I}_+/J$ are undefined and
cannot be used by continuity alone. The appropriate starting point is instead the
uneliminated trial polynomial \eqref{cubic-pushforward}, whose pairing and mixing
functional on $H^2(M_8)=H^2(Y_3)\oplus\RR u_B$ take the form
\begin{equation}
 \mathcal K=\begin{pmatrix}\mathsf{L}_{\mathfrak n}&0\\0&0\end{pmatrix},
 \qquad h=(\mathfrak n^2,0).
 \label{MSW-degenerate-block}
\end{equation}
Hard Lefschetz makes $\mathsf{L}_{\mathfrak n}$ invertible on the base sector since
$-\mathfrak n=P$ is a K\"ahler class.  Writing the trial mixing class as
$\beta+s u_B$, stationarity gives
\begin{equation}
 \beta_*=-\mathfrak n,\qquad
 k_{RR}^{(0)}=\frac13\int_{Y_3}\mathfrak n^3,
 \qquad c^{(0)}=-\int_{Y_3}\mathfrak n^3.
 \label{MSW-base-extremum}
\end{equation}
Although the retained anomaly does not depend on $s$, the null fibre direction leaves
the corresponding trial-current component undetermined, and its boundary treatment
cannot be inferred from this stationarity calculation. In the MSW theory the
supersymmetric R-current is the unit-supercharge Cartan of the enhanced $SU(2)$
R-symmetry. Choosing $s=0$ at the base extremum then gives $\widehat{\mathfrak
G}=\mathfrak n(u_B-r)$, in agreement with the symmetric angular-flux lift on $S^2$.

Since $b=p_1(Y_3)=-2c_2(Y_3)$ and $\ell=0$, the mixed $X_8$ functional
vanishes.  Directly from \eqref{I4-X8-full},
\begin{equation}
 I_4^{X_8}=-\frac1{24}\left(\int_{Y_3}\mathfrak n b\right)r^2
          -\frac1{96}\left(\int_{Y_3}\mathfrak n b\right)p.
 \label{MSW-X8-polynomial}
\end{equation}
Thus, $\delta c_L=\tfrac14\int\mathfrak n b$ and
$\delta c_R=\tfrac12\int\mathfrak n b$.  With
$P^3=\int_{Y_3}P^3$ and $c_2\cdot P=\int_{Y_3}c_2(Y_3)P$, the result is
\begin{equation}
 c_L^{\rm ret}=P^3+\frac12c_2\cdot P,\qquad
 c_R^{\rm ret}=P^3+c_2\cdot P.
 \label{MSW-central-charges}
\end{equation}
With the stated exchange of chiralities and $P^3=6D$, these expressions reproduce
equations (3.14)--(3.15) of \cite{MaldacenaStromingerWitten:1997}. Along $P=NP_0$, they
exhibit the expected $N^3$ and $N$ scaling, thereby checking the cubic, direct
R-symmetry and gravitational coefficients against the known MSW result. The mixed
R--Betti term vanishes in this specialization, so it is not tested by the MSW
comparison. Instead, the example demonstrates that the original anomaly polynomial
continues to yield the correct central charges when a decoupled null block makes the
inverse-block ratio $\mathsf{I}_+/J$ undefined.

\section{Curve products and projective examples}
\label{curves}
\label{applications}

\subsection{Curve-product anomaly formulae}

We now specialize to a product of compact Riemann surfaces $C_i$ of
genera $g_i$,
\begin{equation}
    B_6=C_1\times C_2\times C_3.
\end{equation}
Let
\begin{equation}
    \kappa_i=1-g_i,
\end{equation}
and choose generators $\omega_i\in H^2(C_i)$ normalized by
\begin{equation}
    \int_{C_i}\omega_i=1.
\end{equation}
The projective twisting and fibre-integrated flux are
\begin{equation}
    \ell
    =
    2\sum_{i=1}^3\kappa_i\omega_i,
    \qquad
    \mathfrak n
    =
    \sum_{i=1}^3N_i\omega_i.
\end{equation}

Since each $C_i$ is two-dimensional,
\begin{equation}
    \pone(TB_6)=0.
\end{equation}
Define
\begin{equation}
    S
    =
    N_1\kappa_2\kappa_3
    +
    N_2\kappa_3\kappa_1
    +
    N_3\kappa_1\kappa_2,
    \label{S-def}
\end{equation}
and
\begin{equation}
    K_{\rm top}
    =
    \kappa_1\kappa_2\kappa_3.
    \label{Ktop-def-curves}
\end{equation}
Then
\begin{equation}
    \int_{B_6}\mathfrak n\ell^2=8S,
    \qquad
    \int_{B_6}\ell^3=48K_{\rm top}.
\end{equation}
For the curve products and examples that follow, we use $\delta c_{L,R}$ for the
$C_3\wedge X_8$ contribution evaluated at the classical extremum. This quantity is
homogeneous of degree one in the unshifted flux and excludes both re-expansions of the
leading term $c^{(0)}$ along a different charge family and contributions from other
anomaly sectors. In these conventions, the general projective formulae simplify to
\begin{equation}
    \delta c_L
    =
    -6S
    -
    12K_{\rm top}\frac{\mathsf{I}_+}{J},
\label{curves-cL}
\end{equation}
and
\begin{equation}
    \delta c_R
    =
    -4S
    -
    12K_{\rm top}\frac{\mathsf{I}_+}{J}.
\label{curves-cR}
\end{equation}
The corresponding gravitational anomaly is
\begin{equation}
    \delta(c_L-c_R)=-2S.
\label{curves-anomaly}
\end{equation}

In the reference-current basis \eqref{G-AyB}, a torus factor gives
$K_{\rm top}=0$, so the mixed R--Betti contribution from $X_8$ vanishes.
This vanishing is preserved under reference-current shifts by retained
Betti currents because $X_8$ produces no Betti--Betti anomaly
\eqref{R-betti-X8}.  For nonzero $K_{\rm top}$, the second terms in
\eqref{curves-cL}--\eqref{curves-cR} are generically present. Although
the direct/mixed separation depends on the reference current, their
extremized sum is invariant.

The formulae can now be evaluated on three magnetically flux-quantized backgrounds
studied in \cite{Gregory:nonpoly}. Their retained higher-derivative anomaly contributions depend on the
topology, magnetic fluxes and classical R--Betti extremum, so no
higher-derivative solution of the metric equations is needed for this
characteristic-class calculation.

To keep the meaning of these applications clear, table~\ref{comparison-scope}
distinguishes the different comparisons made in the paper. In particular, the direct
fibre integral and the matrix calculation check the evaluation and extremization of the
specified anomaly polynomial, rather than constituting an independent determination of
the full higher-derivative central charges.
\begin{table}[htbp]
\centering
\small
\renewcommand{\arraystretch}{1.2}
\begin{tabular}{@{}>{\raggedright\arraybackslash}p{0.30\textwidth}p{0.65\textwidth}@{}}
\toprule
Comparison & What it establishes\\
\midrule
Fixed-section sum and fibre integral &
Agreement of two evaluations of the same projective characteristic-class
push-forward with fixed coupling, flux lift and source normalization.\\
Closed twelve-dimensional test &
The mixed characteristic number on the specified closed spin space,
with unit R and fibre Betti sources.\\
Factor/fibre matrix solve &
The classical mixing coefficients and their substitution into the
projective $X_8$ polynomial for the quoted flux vector.\\
Wrapped-M5 field-theory anomaly &
Agreement with independently known anomaly data through orders $N^3$
and $N$ on the $S^4$-bundle topology.\\
Calabi--Yau/MSW specialization &
Agreement of the cubic and linear central charges with the MSW result,
after a chirality exchange.  The mixed R--Betti term vanishes here.\\
Endpoint and flux central charges &
Numerical consistency of two expressions for the same leading endpoint
observable.  Their equality on the endpoint variety is established in
\cite[section~6.3]{Gregory:nonpoly}.\\
\bottomrule
\end{tabular}
\caption{Scope of the anomaly and central-charge comparisons.}
\label{comparison-scope}
\end{table}

\subsection{Flux cycles and background data}

We take the magnetically quantised flux vectors from table~1 of
\cite{Gregory:nonpoly}. As explained in section~4.3 of that paper, these
periods satisfy the shifted magnetic quantization law, but they do not by
themselves fix compact electric Page-charge periods or flat $C$-field data
when such data are present. Determining those Page periods, where present, would also require
the global patching of $C_3$ and possible flat $C$-field data, and is
separate from the magnetic quantization statement. Beyond the numerical
representatives in section~5 and their reconstruction
data in appendix~A.1, that work establishes regular families in the admissible
curvature-sign sectors, proves that non-polynomial profiles are generic within
those families, and certifies a finite-ratio flux neighbourhood in the fully
positive sector (section~4.2 and appendix~C).
The fully positive $(+,+,+)$ representative lies in this certified box,
whereas the $(+,+,0)$ and $(+,0,0)$ finite-ratio representatives are supported
by numerical boundary-value solutions rather than that certification.
The existence and regularity
analysis supplies the geometric input to the present calculation, which
determines the order-$N$ contribution of the specified anomaly couplings.
Appendix~\ref{projective-benchmark} gives an endpoint benchmark for the first
example, while the full factor/fibre matrix calculation below treats the fully
curved background.

To specify the flux vectors without reference to a coordinate ansatz,
orient the curves so that $\int_{C_i}\omega_i=1$, use the product orientation
$\int_{B_6}\omega_1\omega_2\omega_3=1$, and orient the fibre by
$\int_{\mathbb{CP}^1}u_B=1$.  For cyclic $(i,j,k)$, let
\begin{equation}
 F_i=\pi^{-1}(C_i),\qquad D_i^-=s_-(C_j\times C_k),\qquad
 N_i=\int_{F_i}\mathfrak g,\qquad M_i=\int_{D_i^-}\mathfrak g,
\end{equation}
where all other base coordinates are fixed and $s_-$ is the section with
$u_B|_{B_-}=0$.  Thus, writing column vectors $n=(N_1,N_2,N_3)^T$,
$m=(M_1,M_2,M_3)^T$ and $\kappa=(\kappa_1,\kappa_2,\kappa_3)^T$,
\begin{equation}
 \mathfrak n=\sum_iN_i\omega_i,\qquad
 f_-=M_1\omega_2\omega_3+M_2\omega_3\omega_1+M_3\omega_1\omega_2,
 \qquad \ell=2\sum_i\kappa_i\omega_i.
\end{equation}
In these dual bases, multiplication by $\mathfrak n$ and the positive-section
flux are represented by
\begin{equation}
 \mathsf{L}_{\mathfrak n}=\begin{pmatrix}0&N_3&N_2\\N_3&0&N_1\\N_2&N_1&0\end{pmatrix},
 \qquad m_+=m+2\mathsf{L}_{\mathfrak n}\kappa.
\end{equation}
All three examples have $\det\mathsf{L}_{\mathfrak n}=2N_1N_2N_3\ne0$,
and can be reproduced by
\begin{align}
 \mathsf{L}_{\mathfrak n}^{-1}&=\frac{1}{2N_1N_2N_3}
 \begin{pmatrix}
 -N_1^2&N_1N_2&N_1N_3\\
 N_1N_2&-N_2^2&N_2N_3\\
 N_1N_3&N_2N_3&-N_3^2
 \end{pmatrix},\\
 \mathsf{I}_3&=6N_1N_2N_3,\qquad \mathsf{I}_-=n^Tm,\qquad
 \mathsf{I}_+=n^Tm_+,\qquad J=m_+^T\mathsf{L}_{\mathfrak n}^{-1}m.
 \label{example-reproduction}
\end{align}
The shifted integral periods are $N_i$ and $M_i^{\rm q}=M_i-2\kappa_j\kappa_k$, whereas
the formulae above are written in terms of the unshifted $M_i$. For later reference, the
intermediate invariants are
\begin{equation}
\begin{array}{c|r|r|c}
 (\kappa_1,\kappa_2,\kappa_3)& \mathsf{I}_-&\mathsf{I}_+&J\\ \hline
 (1,1,0)&-126&126&17209/165\\
 (1,1,1)&-39918888&39918888&225176/3\\
 (1,0,0)&-7429968&7429968&2572033/567
\end{array}
\end{equation}
Combined with $c^{(0)}=-\mathsf{I}_3+3\mathsf{I}_+\mathsf{I}_-/J$, these invariants allow both the leading central
charges and the corrections in the following examples to be evaluated exactly as
rational numbers.

\subsection{Non-polynomial example with one torus factor \texorpdfstring{$(+,+,0)$}{(+,+,0)}}

The sign triples label the constant-curvature signs of the three curve factors,
with $+$ corresponding to a sphere and $0$ to a torus. The first background
lies in the $(+,+,0)$ chamber, where one curve factor is a torus.
This immediately sets the topological coefficient controlling the mixed term to
\begin{equation}
    K_{\rm top}=0.
\end{equation}
In the canonical ordering adapted to
$B_6=S^2\times S^2\times T^2$, a convenient representative has unshifted
periods
\begin{equation}
    (N_1,N_2,N_3;M_1,M_2,M_3)
    =
    (-10,-11,-3;10,10,-28),
\end{equation}
and shifted integral periods
\begin{equation}
    (-10,-11,-3;10,10,-30).
    \label{pluspluszero-shifted-flux}
\end{equation}
Although the torus factor removes the mixed $X_8$ coefficient, it does not trivialize
the classical geometry. Indeed, $S=N_3=-3$, and the leading central charge obtained by
localization is
\begin{equation}
    c^{(0)}
    =
    \frac{26215200}{17209}
    \simeq
    1523.34243709687.
\end{equation}
Substitution into the curve-product anomaly formulae
\eqref{curves-cL}--\eqref{curves-cR} then determines the linear-flux corrections as
\begin{equation}
    \delta c_L=18,
    \qquad
    \delta c_R=12,
    \qquad
    \delta(c_L-c_R)=6.
    \label{pluspluszero-correction}
\end{equation}
In the reference-current basis \eqref{G-AyB}, the $X_8$ mixed R--Betti coefficient
vanishes despite the non-trivial two-derivative extremization that selects the leading
R-current. The resulting correction is entirely direct in this decomposition, giving
an explicit example of how the topology of the base distinguishes the subleading
anomaly from its classical mixing structure.

\subsection{Fully curved example \texorpdfstring{$(+,+,+)$}{(+,+,+)}}
\label{fully-curved-example}

The second example has three positively curved factors, $B_6=S^2\times S^2\times S^2$,
so the topological invariant is
\begin{equation}
    K_{\rm top}=1,
\end{equation}
and the mixed R--Betti contribution is present. In the canonical cycle basis, the
unshifted flux vector is
\begin{equation}
    (N_1,N_2,N_3;M_1,M_2,M_3)
    =
    (-1098,-3294,-3721;10248,10248,-1368),
    \label{plusplusplus-flux}
\end{equation}
while the shifted integral vector is
\begin{equation}
    (-1098,-3294,-3721;10246,10246,-1370).
    \label{plusplusplus-shifted-flux}
\end{equation}
We evaluate the characteristic polynomial on the unshifted class, consistently with the
conventions established in section~\ref{projective}. The corresponding projective
invariants are
\begin{equation}
    \mathsf{I}_+=39918888,
    \qquad
    J=\frac{225176}{3},
    \qquad
    S=-8113.
    \label{plusplusplus-invariants}
\end{equation}
Substitution into the leading Schur-complement formula yields
\begin{equation}
    c^{(0)}
    =
    \frac{480133225257552}{28147}.
\end{equation}

To check the mixed contribution directly, we use the representatives
\eqref{projective-CW-representatives}, where $Y$ is a closed two-form
representing the projective class $u_B$. In the order
$(\omega_1,\omega_2,\omega_3,Y)$, the fibre integral of
$\widehat{\mathfrak G}^{\,3}/6$ gives
\begin{equation}
 \cK=\begin{pmatrix}
 0&-3721&-3294&-3782\\
 -3721&0&-1098&610\\
 -3294&-1098&0&-10152\\
 -3782&610&-10152&-26648
 \end{pmatrix},\qquad
 h=\begin{pmatrix}24513948\\8171316\\7233624\\79837776\end{pmatrix}.
 \label{plusplusplus-direct-matrix}
\end{equation}
Solving $\cK t_0+h=0$ gives
\begin{equation}
 (t_0)_B=\frac{14969583}{28147}=\frac{\mathsf{I}_+}{J}.
 \label{plusplusplus-direct-mixing}
\end{equation}
The full matrix solution therefore reproduces the projective ratio $\mathsf{I}_+/J$, while
applying the same direct fibre integration to $\widehat{\mathfrak G}X_8$ yields
\begin{equation}
 I_4^{X_8}=-8113r^2+2r x_B+\frac{8113}{12}p,
 \qquad p=\pone(TW_4),
 \label{plusplusplus-direct-X8}
\end{equation}
whose mixed coefficient agrees with the closed test
\eqref{projective-closed-mixed-number}.  Substitution of
$x_B=(t_0)_B r$ changes the signed $k_{RR}$ by $4(t_0)_B$, and hence
changes each central charge by $-12(t_0)_B$.

The direct geometric terms are
\begin{equation}
    \delta c_L^{\rm dir}=48678,
    \qquad
    \delta c_R^{\rm dir}=32452,
\end{equation}
whereas the common R--Betti contribution is
\begin{equation}
    \delta c_{L}^{\rm mix}
    =
    \delta c_{R}^{\rm mix}
    =
    -12\frac{\mathsf{I}_+}{J}
    =
    -\frac{179634996}{28147}.
    \label{plusplusplus-mixing-piece}
\end{equation}
Combining the direct and mixed contributions, we obtain
\begin{equation}
    \delta c_L
    =
    \frac{1190504670}{28147},
    \qquad
    \delta c_R
    =
    \frac{733791448}{28147}.
    \label{plusplusplus-correction}
\end{equation}
Their difference is
\begin{equation}
    \delta(c_L-c_R)=16226=-2S,
\end{equation}
as required by the mixing independence of the gravitational anomaly. Numerically, the
mixed term reduces each direct central-charge correction by approximately
$6.38\times10^3$, making its contribution significant even though it does not affect the
left--right difference.

The nonzero entry in \eqref{plusplusplus-direct-X8} is the $X_8$
contribution to the mixed anomaly.  For the full retained polynomial, set
$m=(0,0,0,2)^T$.  The corrected extremum obeys
\begin{equation}
 t_{\rm ret}=t_0-\cK^{-1}m,\qquad
 h+m+\cK t_{\rm ret}=0.
 \label{plusplusplus-corrected-mixed-zero}
\end{equation}
The vanishing of the total R--Betti anomaly at the corrected extremum is the
stationarity condition. Along an unshifted-flux ray, the displacement changes the
extremized anomaly only at order $N^{-1}$, as explained in
section~\ref{hd-extremization}. Thus, this fully curved example exhibits both a
nonzero order-$N$ mixed contribution and the expected hierarchy of corrections
to the extremal current.

\subsection{Vanishing \texorpdfstring{$X_8$}{X8} contribution \texorpdfstring{$(+,0,0)$}{(+,0,0)}}

The final background is a product with two torus factors, for which
\begin{equation}
    B_6=S^2\times T^2\times T^2,
    \qquad
    \ell=2\omega_1,
\end{equation}
so $\ell^2=0$ and the $p_1/4$ shift vanishes.  The unshifted and shifted
integral period vectors therefore agree,
\begin{equation}
    (N_1,N_2,N_3;M_1,M_2,M_3)
    =
    (-13104,-4536,-819;728,-1664,6640).
    \label{pluszerozero-flux}
\end{equation}
The leading central charge is
\begin{equation}
    c^{(0)}
    =
    \frac{657354424810696704}{2572033}
    \simeq
    2.55577756899\times10^{11}.
\end{equation}
At the same time,
\begin{equation}
    S=0,
    \qquad
    K_{\rm top}=0.
\end{equation}
Thus, the $C_3\wedge X_8$ characteristic-class contribution vanishes,
\begin{equation}
    \delta c_L
    =
    \delta c_R
    =
    0.
    \label{pluszerozero-correction}
\end{equation}
The example therefore shows that a non-trivial holographic fixed point can have no
linear-$N$ correction from $C_3\wedge X_8$, with the vanishing controlled here by
topology rather than by the detailed metric functions.

Across all three examples, the differences of the quoted corrections agree with
$\delta(c_L-c_R)=-2S$, as required by the gravitational term in the anomaly polynomial.
The common shifts must still be obtained from the full retained trial polynomial,
since they cancel in this difference. The agreement therefore supplies a uniform,
mixing-independent check of the relative central-charge corrections in all three
curvature-sign sectors.

\section{Discussion}

The central contribution of this work is the projective trial-current anomaly,
including the previously unavailable mixed R--Betti coefficient from $C_3\wedge X_8$.
At leading order, the Betti intersection pairing also explains how the invariant
$J$ arises as the Schur-complement denominator in the central-charge formula
previously obtained by localization \cite{GregoryNotmanSparks:2026localizing},
\begin{equation}
 c^{(0)}=-\mathsf{I}_3+3\frac{\mathsf{I}_+\mathsf{I}_-}{J}.
\end{equation}
The projective $X_8$ calculation then combines this new mixed coefficient with
the direct R-symmetry term to yield
\eqref{cL1-master}--\eqref{cR1-master}, while the gravitational anomaly
\eqref{grav-anomaly} remains independent of mixing and of the projective quantization
shift.

Equality of the Pontryagin-class restrictions removes the section flux
from the $X_8$ trial polynomial, although it still enters the classical
mixing parameters.  When $\ell=c_1(TB_6)$, the mixed coefficient in
the chosen fibre Betti basis is $2\chi(B_6,\cO_{B_6})$.
The resulting index explains why the mixed coefficient vanishes for curve products
containing a torus but is nonzero for $(S^2)^3$. In the latter case its contribution
must be added to the direct term at the extremum before drawing a
reference-current-independent conclusion, a distinction illustrated explicitly by the
solution data of \cite{Gregory:nonpoly}.

The different comparisons serve complementary purposes. The projective
fixed-section, fibre-integral and closed-space evaluations check the same
characteristic-class coupling and do not independently fix its absolute
normalization. By contrast, the wrapped-M5 construction tests the shared
coefficients against independently known field-theory anomalies on a
different topology, while the MSW specialization reproduces known cubic
and linear central charges
even though a decoupled null direction prevents direct use of the inverse-block formula.
Their respective scopes are summarized in table~\ref{comparison-scope}.

The order-$N$ expressions are defined along an unshifted-flux ray and must be
re-expanded with care along any alternative charge family so that the source-free
constraint remains satisfied. Their identification with full CFT corrections also
depends on the boundary-current assumptions in section~\ref{trial-space-completeness}
and the omitted-sector estimates discussed in section~\ref{omitted-anomaly-sectors}. In
particular, the vanishing found for the $(+,0,0)$ background is a property of the
specified $X_8$ contribution and need not extend to the complete CFT correction.

A further direction would be to construct the Euclidean filling and boundary
representative considered conditionally in appendix~\ref{relative}, thereby providing a
global relative interpretation of the anomaly data without altering the compact
calculation. The method also invites applications to internal spaces with additional
fixed components or larger trial symmetry sectors, where the characteristic-class
push-forward and extremization can again be analysed as separate steps.

\appendix

\section{Technical details of the projective anomaly reduction}
\label{projective-technical}

We collect here the cohomological and characteristic-class steps underlying the compact
formulae, used to support the structure of the trial anomaly
and its extremization. The details also make clear which statements depend on the choice
of equivariant representatives and which are intrinsic to the projective topology.

\subsection{Internal bilinears and the supersymmetry lift}
\label{internal-lift-details}

We use unit-radius $\AdS_3$ and the unwarped internal metric in
\eqref{lorentzian-input}. Let $\epsilon$ be the normalized internal spinor.
With real symmetric eight-dimensional gamma matrices, set $\gamma_9=\gamma_1\cdots\gamma_8$,
$\bar\epsilon=\epsilon^\dagger$, $\epsilon^c=\epsilon^*$ and
$\bar\epsilon\epsilon=1$, $\epsilon^T\epsilon=0$.  We use the subclass
$\epsilon^T\gamma_9\epsilon=0$ and
$\epsilon^T\gamma_{(1)}\epsilon=0$ discussed in
\cite[section~4.3]{Ashmore:2023}, and write
$\gamma_{(p)}=\frac1{p!}\gamma_{m_1\cdots m_p}
 e^{m_1}\wedge\cdots\wedge e^{m_p}$, with unit-weight
antisymmetrization.  Our internal conventions are
\cite[section~4]{BenettiGenolini:2023ads}
\begin{equation}
 \zeta=\bar\epsilon\gamma_9\epsilon,\qquad
 \kappa=\bar\epsilon\gamma_{(1)}\epsilon,\qquad
 L=-i\bar\epsilon\gamma_9\gamma_{(1)}\epsilon,\qquad
 \omega=-i\bar\epsilon\gamma_9\gamma_{(2)}\epsilon,
 \qquad \xi^\flat=\frac12L.
 \label{internal-bilinear-definitions}
\end{equation}
Here $\xi^\flat$ denotes the metric-dual one-form of the vector $\xi$.
The spinorial Lie derivative is $\mathcal L_\xi\epsilon=i\epsilon/2$, so the generator
assigning unit charge to the preserved supercharge is $2\xi$. This accounts for the
shift $\ell\mapsto\ell+2r$ used in the anomaly calculation.
Here $L$ is the spinorial one-form, distinct from the twisting line bundle
$\cL$, while $\kappa$ is unrelated to the curvature parameters $\kappa_i$. We introduce the electric potential and corrected moment function by
\begin{equation}
 \dd a_0=e^{3\lambda}f_{\rm el},\qquad
 \mu=\frac12\left(e^{3\lambda}\zeta-a_0\right).
 \label{electric-moment-potential}
\end{equation}
Our $\mu$ is the function denoted $y$ in
\cite[section~2.4]{GregoryNotmanSparks:2026localizing}, distinct from the
projective cohomology generator $u_B$. The scalar bilinear identity is
$\dd(e^{3\lambda}\zeta-a_0)=2e^{3\lambda}\kappa$.
These potentials are globally defined for the projective geometries considered
here. Indeed, the simply connected sphere fibre gives an isomorphism
$\pi^*\colon H^1(B_6;\RR)\to H^1(M_8;\RR)$, while the closed one-form
$e^{3\lambda}\kappa$ restricts to zero on either fixed section. Its
cohomology class therefore vanishes. The scalar bilinear identity then gives
$[e^{3\lambda}f_{\rm el}]=-2[e^{3\lambda}\kappa]=0$, establishing global
choices of both $a_0$ and $\mu$, also when $H^1(B_6;\RR)$ is nonzero
\cite[section~4.1]{GregoryNotmanSparks:2026localizing}.
 The bilinear identities then yield
\begin{equation}
 \iota_\xi\omega=\frac12\kappa,\qquad
 \dd\mu=e^{3\lambda}\kappa
       =2\iota_\xi(e^{3\lambda}\omega).
 \label{moment-map-differential}
\end{equation}
With the unit-parameter Cartan differential
$\dd_\xi=\dd-\iota_\xi$, where $\iota_\xi$ denotes contraction
with the vector $\xi$, the magnetic polyform is
\begin{equation}
 \Phi^F=e^{3\lambda}F-\frac12e^{3\lambda}\omega-\frac{\mu}{4},
 \qquad \dd_\xi\Phi^F=0.
 \label{AdS3-flux-polyform}
\end{equation}
Its closure follows from the magnetic Bianchi identity together with
$\dd(e^{3\lambda}\omega)=-2\iota_\xi(e^{3\lambda}F)$.
Explicitly,
\begin{align}
 [\dd_\xi\Phi^F]_5&=\dd(e^{3\lambda}F)=0,\\
 [\dd_\xi\Phi^F]_3&=-\tfrac12\dd(e^{3\lambda}\omega)
                       -\iota_\xi(e^{3\lambda}F)=0,\\
 [\dd_\xi\Phi^F]_1&=-\tfrac14\dd\mu
                       +\tfrac12\iota_\xi(e^{3\lambda}\omega)=0.
 \label{closure-components}
\end{align}
All the forms are $\xi$-invariant, and the displayed formulae use a Cartan parameter set
to one. Restoring a formal degree-two parameter $u$ replaces the equivariant
differential by $\dd-u\iota_\xi$ and multiplies the components of degrees $(4,2,0)$ by
$(1,u,u^2)$. The additive freedom in $a_0$ induces a constant shift of $\mu$ and hence
an external degree-four shift of the homogenized flux lift. This freedom is
accounted for by the flux-lift analysis of section~\ref{flux-lift-ambiguities}.
Thus, the globally closed polyform supplies a supersymmetry reference lift
for backgrounds with electric as well as magnetic flux. In the purely
magnetic subclass, choosing $a_0=0$ recovers
$\mu=\tfrac12e^{3\lambda}\zeta$.

\subsection{Equivariant flux lifts and reference-current independence}
\label{flux-lift-ambiguities-details}

The anomaly reduction depends on an equivariant extension of the magnetic flux and not
merely on its ordinary cohomology class. We therefore explain the choice
\eqref{G-AyB} and relate it to the supersymmetry lift of
appendix~\ref{internal-lift-details}. General equivariant flux extensions
 and their
ambiguities are discussed in \cite{Bah:2019inflow}.  The projective ring
makes the argument explicit here.  We work in real cohomology and with local
anomaly polynomials, leaving torsion and global anomalies outside the scope
of the calculation.

For the fibre action, with the physical normalization $u_\xi=2r$, the
Borel equivariant cohomology ring is
\begin{equation}
 H^*_{U(1)_R}(M_8;\RR)
 =H^*(B_6;\RR)[r,u_B]/\bigl(u_B(u_B-\ell-2r)\bigr).
 \label{lift-equivariant-ring}
\end{equation}
The base is fixed by this action.  The projective bundle formula makes this
a free module with basis $1,u_B$ over $H^*(B_6;\RR)[r]$.
Thus, an equivariant degree-four class restricting to
$\mathfrak g=f_-+u_B\mathfrak n$ at $r=0$ has the form
\begin{equation}
 \widehat{\mathfrak g}'=
 f_-+u_B\mathfrak n+r\,\widehat\lambda+\nu r^2,
 \qquad
 \widehat\lambda=\lambda_B+s_\lambda u_B,
 \quad \lambda_B\in H^2(B_6;\RR),\quad s_\lambda,\nu\in\RR.
 \label{lift-general-difference}
\end{equation}
This is a statement in equivariant cohomology, so changing a representative by an
equivariantly exact form does not affect its push-forward over the closed manifold. In
particular, the globally defined internal supersymmetry polyform, after homogenization
and normalization by $(2\pi\ellp)^3$, lifts the same magnetic class $\mathfrak g$ and
can differ from the chosen reference by precisely the terms displayed above. Equality of
their fixed-section representatives term by term is not required. In
particular, including the electric potential in $\mu$ changes the
supersymmetry representative without changing $\mathfrak g$ or the
projective ring. The corrected lift is therefore covered by the same
reference-current and external-four-form ambiguities, provided the relevant
Betti directions belong to the retained trial sector.

Keep a fixed basis of Betti currents, with equivariant extensions
$\widehat\eta_I=\beta_I+s_Iu_B$.  If
$\lambda=v^I\eta_I$ belongs to the retained trial sector, then
\begin{equation}
 \widehat{\mathfrak G}'(r,x)
 =\widehat{\mathfrak G}(r,x+v r)+\gamma_4.
 \label{lift-source-redefinition}
\end{equation}
Here $\gamma_4$ is purely external and of degree four. It includes
$\nu r^2$ and the terms $r x^I$ induced by changing the equivariant lifts
of the fixed $\eta_I$.  A purely external gravitational term can also be
included.  The shift $x\mapsto x+v r$ is a change of reference R-current by
Betti currents. Since those currents leave the supercharge neutral, its
unit R-charge is preserved.  The argument requires the relevant Betti
directions to be retained and does not replace the physical trial-space
analysis of section~\ref{trial-space-completeness}.  Non-integral
basis changes are allowed for continuous trial extremization but do not
identify the integral charge lattice.

To evaluate the remaining ambiguity, use the same sign convention as in
\eqref{bulk-boundary-anomaly-dictionary} and define the retained local
anomaly polynomial by
\begin{equation}
 I_4[\widehat{\mathfrak G}]
 =\left[\int_{M_8}
 \left(\frac16\widehat{\mathfrak G}^{\,3}
       +\widehat{\mathfrak G}\,\widehat X_8\right)\right]_4.
 \label{lift-local-anomaly}
\end{equation}
External-degree counting gives
\begin{equation}
 I_4[\widehat{\mathfrak G}+\gamma_4]
 -I_4[\widehat{\mathfrak G}]
 =\gamma_4\int_{M_8}
 \left(\frac12\mathfrak g^2+X_8(TM_8)\right).
 \label{lift-gamma-variation}
\end{equation}
Only the purely internal pieces can multiply $\gamma_4$ in this
four-form anomaly.  On the ungauged projective bundle, the stable tangent
splitting gives
\begin{equation}
 p_1(TM_8)=\pi^*\bigl(p_1(B_6)+\ell^2\bigr),\qquad
 p_2(TM_8)=\pi^*\bigl(p_2(B_6)+p_1(B_6)\ell^2\bigr).
\end{equation}
Consequently, $X_8(TM_8)$ is pulled back from a degree-eight class on the
six-dimensional base and vanishes in real cohomology.  Its equivariant extension nevertheless
has external-degree components that produce the anomalies below.
Independently,
\begin{equation}
 \int_{M_8}\mathfrak g^2
 =\int_{B_6}\bigl(2f_-\mathfrak n+\ell\mathfrak n^2\bigr)
 =\mathsf{I}_-+\mathsf{I}_+=0.
 \label{lift-tadpole-projective}
\end{equation}
For smooth compact source-free two-derivative backgrounds, the last
equality follows directly from the internal degree-eight component of the
full $C_3$ equation of motion,
\begin{equation}
 \dd\left(e^{6\lambda}*_8f_{\rm el}\right)
 +\frac12e^{6\lambda}F\wedge F=0,
 \label{electric-integrated-tadpole}
\end{equation}
where $*_8$ is defined using the unwarped internal metric and the
orientation of \cite[section~2]{GregoryNotmanSparks:2026localizing}.
Integration over the closed $M_8$ gives $\int_{M_8}\mathfrak g^2=0$
and hence \eqref{primitivity}, without requiring pointwise vanishing of
$F\wedge F$. The retained $X_8$ term has zero integrated internal class
on this projective topology, as established above. Thus, the integrated
constraint also makes \eqref{lift-gamma-variation} vanish in the retained
anomaly calculation.
This constrains the admissible background flux rather than the trial
parameters.  The same conclusion need not hold for arbitrary projective
fluxes violating this condition or in the presence of additional sources.

It remains to check reference-current independence explicitly. Write the
gauge part of the retained anomaly as
\begin{equation}
 I_{4,\mathrm{gauge}}
 =\frac12 k\,r^2+r\,b^T x+\frac12x^T\cK x.
\end{equation}
Under \eqref{lift-source-redefinition}, after discarding the vanishing
$\gamma_4$ contribution, its entries transform as
\begin{equation}
 k'=k+2b^Tv+v^T\cK v,\qquad
 b'=b+\cK v,\qquad \cK'=\cK.
 \label{lift-anomaly-basis-change}
\end{equation}
For nondegenerate $\cK$ this gives
\begin{equation}
 k'-b'^{T}\cK^{-1}b'=k-b^T\cK^{-1}b,
 \qquad t_*'=t_*-v.
 \label{lift-schur-invariance}
\end{equation}
The gravitational coefficient is likewise unaffected, and the identity holds for the
combined cubic and $X_8$ polynomial. Consequently, it also protects the order-$N$ result
obtained at the classical extremum, even though the separate direct and mixed
contributions vary with the reference current. Only their extremized sum is invariant.

Outside the retained Betti sector and the source-free setting, equality of
different lifts would require additional current-sector information, as discussed
in section~\ref{trial-space-completeness}. The independent global filling
condition is stated in appendix~\ref{global-euclidean-scope}. Within the hypotheses used in
the main text, however, the exact Schur-complement identity proves that the
reference lift suffices for the extremized anomaly without an explicit determination
of $v^I$ or $\nu$.

\subsection{The gauged projective characteristic-class extension}

Let
\begin{equation}
    r=c_1(\cR)
\end{equation}
denote the extension of the two-dimensional R-symmetry background.  In the
normalization in which the supercharge has unit R-charge, the projective
fibre rotation is gauged through
\begin{equation}
    \ell\longrightarrow\ell+2r.
    \label{R-twist-details}
\end{equation}
A natural characteristic-class filling is therefore
\begin{equation}
    W_{12}^{\rm top}
    =P\!\left(\cO\oplus\cL\boxtimes\cR^2\right)
    \longrightarrow B_6\times W_4,
    \qquad \partial W_4=M_3.
\label{W12}
\end{equation}
The superscript ``top'' distinguishes this characteristic-class extension
from a supersymmetric relative filling.

The vertical tangent line obeys
\begin{equation}
    c_1(T_\pi)=2u_B-(\ell+2r),
\end{equation}
and the gauged projective relation implies
\begin{equation}
\pone(T_\pi)=(\ell+2r)^2.
\label{p1-vertical-R-details}
\end{equation}
The spin condition can also be checked directly using the second
Stiefel--Whitney class $w_2$. Modulo two,
\begin{equation}
    w_2(T_\pi)=\ell.
\end{equation}
In the projective backgrounds of \cite{Gauntlett:2006ns}, the relation $\ell\equiv
c_1(B_6)\pmod 2$ identifies $\ell$ with $w_2(B_6)$ modulo two. If $W_4$ is chosen spin,
this implies
\begin{equation}
    w_2(W_{12}^{\rm top})=0.
\end{equation}
Thus, the natural gauged extension has the required spin topology.

Because \eqref{p1-vertical-R-details} is independent of $u_B$, so is $X_8$.  The
unshifted magnetic class then obeys the exact push-forward identity
\begin{equation}
    \pi_*\bigl(\widehat{\mathfrak g}\,X_8\bigr)
    =\widehat{\mathfrak n}\,X_8.
\label{pushforward-gX8-details}
\end{equation}
In particular, all dependence on the section flux $f_-$ cancels before any
specialization of the base.

\subsection{Direct fibre integration and the closed projective test}
\label{projective-direct-check-details}

A direct Chern--Weil calculation provides another way to obtain the mixed coefficient,
without invoking a fixed-section sum. Put $a=\ell+2r$, and on the complement of the pole
sections choose a normalized angular form $\vartheta$ together with a fibre moment
coordinate $\rho\in[0,1]$ satisfying
\begin{equation}
 \dd\vartheta=a,\qquad \int_{S^1}\vartheta=1,
 \qquad \rho|_{B_-}=0,\qquad \rho|_{B_+}=1.
\end{equation}
The closed two-forms
\begin{equation}
 Y=\rho a+\dd\rho\wedge\vartheta,\qquad
 v=2Y-a=(2\rho-1)a+2\dd\rho\wedge\vartheta
 \label{projective-CW-representatives}
\end{equation}
extend smoothly across the poles and represent $u_B$ and $c_1(T_\pi)$, respectively, with
the fibre orientation fixed by $\int_{\mathbb{CP}^1}Y=1$. Closed characteristic numbers
can be evaluated using a connection adapted to the stable tangent splitting, since a
change of connection alters the characteristic forms only by exact terms.

Since $(\dd\rho\wedge\vartheta)^2=0$, ordinary fibre integration gives
\begin{align}
 \pi_*(v^{2j})
 &=4j\,a^{2j-1}\int_0^1(2\rho-1)^{2j-1}\dd\rho=0,
 &&j\geq1,\label{projective-CW-even}\\
 \pi_*(Yv^{2j})
 &=a^{2j}\int_0^1\frac{\dd}{\dd\rho}
       \left[\rho(2\rho-1)^{2j}\right]\dd\rho
 =a^{2j},
 &&j\geq0.\label{projective-CW-Y-even}
\end{align}
These fibre-integral identities hold for the explicit representatives even though the
projective relation $Y(Y-a)=0$ need not hold pointwise as an equality of differential
forms.

For the mixed R--Betti term, external Pontryagin classes cannot contribute
by degree.  With $b=\pone(TB_6)$, the relevant part of $X_8$ is
\begin{equation}
 \frac1{192}\left[b^2-4\ptwo(TB_6)-2bv^2+v^4\right].
\end{equation}
For $\widehat\eta=\beta+sY$, equations
\eqref{projective-CW-even}--\eqref{projective-CW-Y-even} remove the
$\beta$ term and replace $Yv^{2j}$ by $a^{2j}$.  Selecting the term linear
in $r$ therefore gives
\begin{equation}
 \delta k_{R\eta}^{X_8}
 =\left.\frac{s}{192}\int_{B_6}
       \frac{\partial}{\partial r}\left(a^4-2ba^2\right)\right|_{r=0}
 =\frac{s}{24}\int_{B_6}\ell(\ell^2-b).
 \label{R-betti-CW-check}
\end{equation}
This reproduces \eqref{R-betti-X8} without a fixed-section sum.

A closed characteristic-number test isolates the nonzero coefficient on
the topology of the fully curved example.  Take
\begin{equation}
 \begin{gathered}
 B_6=(\mathbb{CP}^1)^3,\qquad
 \ell=2(\omega_1+\omega_2+\omega_3),\qquad
 \int_{B_6}\omega_1\omega_2\omega_3=1,\\
 W_4=\mathbb{CP}^1_R\times\mathbb{CP}^1_B,\qquad
 r^2=x_B^2=0,\qquad \int_{W_4}r x_B=1,
 \end{gathered}
\end{equation}
where $x_B$ is the unit background class for the Betti current associated
with $u_B$.  The total space
\begin{equation}
 Z_{12}=P\!\left(\cO\oplus\cL\boxtimes\cR^2\right)
       \longrightarrow B_6\times W_4
 \label{projective-closed-test-space}
\end{equation}
is a smooth closed spin manifold.  Its restriction over a point of $W_4$
is the original $M_8$.  The base Pontryagin classes vanish, so the splitting
connection gives $X_8=v^4/192$.  The response of the $C_3X_8$ characteristic
number to the source insertion $x_BY$ is
\begin{equation}
 \delta k_{RB}^{X_8}
 =\int_{Z_{12}}x_BY X_8
 =\frac1{192}\int_{B_6\times W_4}x_B(\ell+2r)^4
 =\frac{2^4\,4!}{192}=2.
 \label{projective-closed-mixed-number-details}
\end{equation}
Here the pure gauge squares and the external gravitational class vanish,
so the test measures precisely the mixed coefficient in the stated
orientation.  It uses a closed characteristic-number evaluation and
requires no supersymmetric filling.

For a smooth projective threefold base satisfying $\ell=c_1(TB_6)$, the
same coefficient admits an index interpretation.  Write
$c_i=c_i(T^{1,0}B_6)$.  Using $\pone(TB_6)=c_1^2-2c_2$ and
Hirzebruch--Riemann--Roch \cite{Stacks:RiemannRoch}, we obtain
\begin{equation}
 \delta k_{R\eta}^{X_8}
 =\frac{s}{12}\int_{B_6}c_1c_2
 =2s\,\chi(B_6,\cO_{B_6}),\qquad
 \chi(B_6,\cO_{B_6})=\sum_{q=0}^3(-1)^q h^{0,q}(B_6).
 \label{R-betti-holomorphic-index-details}
\end{equation}
For a product of curves this is $2s\prod_i(1-g_i)$, giving $2s$ for a
product of three spheres and zero whenever a torus factor is present.  This is the
holomorphic Euler characteristic, not the topological Euler number.
Interpreting this coefficient as a boundary anomaly still requires the
current-sector assumptions used in extremization. As a characteristic-class
calculation, the direct integral independently checks the projective $C_3X_8$
coefficient and identifies its mixed contribution with a holomorphic index.

\section{Subleading extremization and current-sector checks}
\label{extremization-details}

In the main text, we retain the charge-family conventions and the
sector-specific large-$N$ formula \eqref{Schur-largeN-retained}. Here we
record the longer charge-ray argument, the general block expansion and the
additional current-sector checks for the curve-product examples.

\subsection{Shifted-charge rays and transverse extensions}
\label{charge-rays-details}

The distinction between an unshifted-flux ray and a ray of shifted integral
charges can be made explicit using \eqref{shifted-charge-ray} and
\eqref{shifted-ray-compatibility}.

For a full shifted-charge ray to obey the source-free constraint, both coefficients must
vanish separately, whereas agreement at a single value of $N$ is insufficient. As an
illustration, scaling the shifted vector $(-10,-11,-3;10,10,-30)$ in
section~\ref{applications} produces unshifted periods $(-10N,-11N,-3N;10N,10N,-30N+2)$
and $\int_{M_8}\mathfrak g(N)^2=12N(N-1)$. The corresponding shifted ray through the
fully curved example gives $32452N(N-1)$, so neither family remains source-free for
positive integers $N>1$. The mixed coefficient in
\eqref{shifted-ray-compatibility} also satisfies
\begin{equation}
 \int_{M_8}\mathfrak q_0\Delta
 =\frac14\int_{B_6}(\pi_*\mathfrak q_0)
       \bigl[p_1(B_6)+\ell^2\bigr].
\end{equation}
Thus, an admissible shifted ray of this form necessarily has vanishing
retained gravitational anomaly.  This restriction does not apply to the
affine charge family \eqref{physical-affine-charge-family}.

For a smooth homogeneous degree-three function $F$ defined near
$\mathfrak q_0$, let $D_\Delta$ denote differentiation in the flux
direction $\Delta$. One may then formally write
\begin{equation}
 F(N\mathfrak q_0+\Delta)
 =N^3F(\mathfrak q_0)
 +N^2D_\Delta F(\mathfrak q_0)
 +\frac N2D_\Delta^2F(\mathfrak q_0)+O(N^0).
 \label{affine-flux-reexpansion}
\end{equation}
Its physical interpretation requires an admissible charge family and a
specified function away from the constraint locus.  In particular, before
imposing $\mathsf{I}_++\mathsf{I}_-=0$, the reference-lift cubic Schur complement is
\begin{equation}
 F_{\rm Schur}=-\mathsf{I}_3-3\frac{\mathsf{I}_+^2}{J},\qquad
 F_{\rm phys}=-\mathsf{I}_3+3\frac{\mathsf{I}_+\mathsf{I}_-}{J}.
 \label{constrained-leading-functions}
\end{equation}
Although the two expressions agree on the admissible locus, they differ away from it by
$F_{\rm Schur}-F_{\rm phys}=-3\mathsf{I}_+(\mathsf{I}_++\mathsf{I}_-)/J$, so their derivatives transverse to that
locus can be different. The reference-lift independence argument of
appendix~\ref{flux-lift-ambiguities-details} also relies on the source-free constraint,
and consequently an arbitrary off-locus continuation of $F_{\rm phys}$ cannot be
interpreted as a physical central charge. To avoid this ambiguity, we perform
re-expansions only along admissible families with nonsingular anomaly blocks, using
\eqref{unshifted-flux-ray} and imposing shifted quantization through
\eqref{physical-affine-charge-family}.

\subsection{General subleading Schur-complement expansion}
\label{general-Schur-expansion}

Returning to the unshifted ray \eqref{unshifted-flux-ray}, we first allow
a general subleading correction to each block within the following local
large-flux hierarchy,
\begin{equation}
\begin{aligned}
 \mathbb A&=N^3\mathbb A_3+N\mathbb A_1+O(N^{-1}),\\
 \mathbb B&=N^2\mathbb B_2+\mathbb B_0+O(N^{-2}),\\
 \mathcal K&=N\mathcal K_1+N^{-1}\mathcal K_{-1}+O(N^{-3}).
\end{aligned}
\label{block-largeN-scaling}
\end{equation}
We retain the coefficient $\mathcal K_{-1}$ to examine the most general algebraic
expansion at this order, without implying that the particular $C_3\wedge X_8$ coupling
generates such a term. The current basis is fixed independently of $N$, and the leading
block $\mathcal K_1$ is assumed nondegenerate. These assumptions define the local matrix
problem and do not cover arbitrary quantum or decoupled sectors.

The inverse Betti block is
\begin{equation}
 \mathcal K^{-1}
 =N^{-1}\mathcal K_1^{-1}
 -N^{-3}\mathcal K_1^{-1}\mathcal K_{-1}\mathcal K_1^{-1}
 +O(N^{-5}).
 \label{Betti-inverse-largeN}
\end{equation}
Although the second term is suppressed in the inverse, multiplication by
two factors of $\mathbb B\sim N^2$ promotes it to order $N$.
Consequently, expanding \eqref{general-Schur} gives
\begin{equation}
\begin{aligned}
 \mathbb A_{\rm eff}
 ={}&N^3\left(
 \mathbb A_3-\mathbb B_2\mathcal K_1^{-1}\mathbb B_2^T\right)\\
 &+N\Bigl[
 \mathbb A_1-\mathbb B_0\mathcal K_1^{-1}\mathbb B_2^T
 -\mathbb B_2\mathcal K_1^{-1}\mathbb B_0^T\\
 &\hspace{18mm}
 +\mathbb B_2\mathcal K_1^{-1}\mathcal K_{-1}
                    \mathcal K_1^{-1}\mathbb B_2^T\Bigr]
 +O(N^{-1}).
\end{aligned}
\label{Schur-largeN}
\end{equation}
It follows that a possible $O(N^{-1})$ term in $\mathcal K$ cannot simply be ignored
when determining the effective anomaly at order $N$. By contrast, the term quadratic in
$\mathbb B_0$, namely $-N^{-1}\mathbb B_0\mathcal K_1^{-1}\mathbb B_0^T$, first
contributes at order $N^{-1}$ and is only one among several terms at that subleading
order in the general hierarchy.

The order-$N$ term may still be evaluated at the classical extremum even
when $\mathcal K_{-1}$ is retained.  To see this without assuming that an
individual leading mixing coefficient is nonzero, set $x=N\tau$ at
fixed geometric source $u$.  Then
\begin{equation}
 \begin{aligned}
 \tau_*&=\tau_0+N^{-2}\tau_1+O(N^{-4}),
 &\tau_0&=-\mathcal K_1^{-1}\mathbb B_2^T u,\\
 \tau_1&=-\mathcal K_1^{-1}
       \left(\mathbb B_0^T u+\mathcal K_{-1}\tau_0\right).
 \end{aligned}
 \label{mixing-shift-general}
\end{equation}
Alongside the pure geometric and mixed terms, the subleading trial polynomial contains
$\frac12\tau^T\mathcal K_{-1}\tau$, whose value at $\tau_0$ accounts for the final
order-$N$ contribution in \eqref{Schur-largeN}. Since the leading polynomial is
stationary there, the linear effect of the displacement of $\tau_*$ cancels, and that
displacement alters the extremized anomaly only at order $N^{-1}$. In the retained
projective sector $\mathcal K_{-1}=0$, reducing the same reasoning to
\eqref{Schur-largeN-retained}.

\subsection{Detailed current-sector checks for products of curves}
\label{trial-current-checks}

Here we detail the boundary prescription and bulk current checks
used in section~\ref{trial-space-completeness}.  For the retained
Chern--Simons gauge fields, diagonalizing the nonzero level matrix allows
a boundary term that fixes one chiral component as a nondynamical source, with the choice of chirality
set by the signed level and orientation.  The conjugate component determines
the current expectation value.  In a Maxwell--Chern--Simons description it
is the flat sector that supplies this conserved current.  The propagating
mode is a separate operator \cite[section~4]{Benini:2013cda}.  The trial
sources used here refer to these global currents, not to boundary gauge
fields integrated over in a different boundary theory.  We assume a
supersymmetric realization of this boundary prescription as specified in
section~\ref{trial-space-completeness}.  The bulk pairing alone does not
settle the treatment of zero-level modes.

Within this current sector, candidate enlargements arise from harmonic
two-forms and geometric isometries.  We can test their coupling to the
R-current for the product backgrounds used below.
For
\begin{equation}
    B_6=C_1\times C_2\times C_3,
\end{equation}
K\"unneth gives
\begin{equation}
    H^2(B_6;\RR)=\bigoplus_{i=1}^3\RR[\omega_i]
    \oplus\bigoplus_{i<j}H^1(C_i;\RR)\otimes H^1(C_j;\RR).
    \label{H2-Kunneth-curves}
\end{equation}
The background classes $\ell$, $\mathfrak n$ and $f_\pm$ lie in the
even subring generated by the $\omega_i$.  Hence any cross-term class
$\beta_\perp$ in the second summand satisfies
\begin{equation}
    \int_{B_6}\beta_\perp\wedge\mathfrak n^2=0,
    \qquad s(\beta_\perp)=0,
    \label{cross-h-vanish}
\end{equation}
and, by the same K\"unneth-degree argument,
\begin{equation}
    \mathcal K(\beta_\perp,\beta_{\rm fac})=0,
    \qquad \mathcal K(\beta_\perp,u_B)=0.
    \label{cross-K-decouple}
\end{equation}
The $X_8$ coupling is also zero on these classes because $s(\beta_\perp)=0$, so they
form a block that decouples from the R/factor/fibre sector through the retained order.
This algebraic decoupling does not necessarily imply uniqueness of the extremum.
Let $\mathcal K_\perp$ be the restriction of the pairing to these cross-term
classes and $t_\perp$ their mixing coefficients. Stationarity requires
\begin{equation}
 \mathcal K_\perp t_\perp=0.
 \label{cross-stationarity}
\end{equation}
If $\mathcal K_\perp$ is invertible, stationarity fixes $t_\perp=0$ uniquely. A kernel
instead leaves flat stationary directions along which the decoupled quadratic anomaly
remains constant. Although the value of the R-anomaly would still be unaffected at this
order, the R-current would not be uniquely selected, and zero-level bulk modes could not
automatically be regarded as boundary global currents without additional infrared and
boundary information.

Here the omitted blocks can be checked explicitly.  Put
$V_{ij}=H^1(C_i;\RR)\otimes H^1(C_j;\RR)$, $i<j$, and let $k$ be the
remaining index.  Write
$\Omega_i(a,a')=\int_{C_i}a\wedge a'$ for the nondegenerate intersection
pairing on $H^1(C_i)$.  The product orientation gives
\begin{equation}
 \mathcal K(a\wedge b,a'\wedge b')
 =-N_k\,\Omega_i(a,a')\Omega_j(b,b'),
 \qquad a,a'\in H^1(C_i),\quad b,b'\in H^1(C_j).
 \label{cross-block-pairing}
\end{equation}
By degree, the spaces $V_{ij}$ are mutually orthogonal, and each nonempty block of
dimension $4g_i g_j$ is nondegenerate precisely when the complementary flux period $N_k$
is nonzero. In integral symplectic bases the corresponding matrices take the form
\begin{equation}
 \mathcal K_\perp|_{V_{ij}}=-N_k\,\Omega_i\otimes\Omega_j,
 \qquad \det(\mathcal K_\perp|_{V_{ij}})=N_k^{4g_i g_j}.
 \label{cross-block-determinant}
\end{equation}
There are no such blocks in the $(+,+,0)$ and $(+,+,+)$ examples, since
at most one factor has nonzero $H^1$.  For $(+,0,0)$ the only extra block
is $V_{23}$, of dimension four, and $N_1=-13104$ makes it nondegenerate.
Thus, the vanishing extra mixing coefficients are unique in these examples.

The retained factor/fibre block is also nondegenerate.  In the cycle basis
specified in section~\ref{applications},
$\det\mathsf{L}_{\mathfrak n}=2N_1N_2N_3$ and
$\det\mathcal K_{\rm fac,fib}=-J\det\mathsf{L}_{\mathfrak n}$.
All three flux vectors have $N_1N_2N_3\ne0$, and their respective values
$J=17209/165$, $225176/3$ and $2572033/567$ are nonzero.  Combined with
\eqref{cross-block-determinant}, this verifies nondegeneracy of the
entire harmonic two-form pairing for the three examples.

For geometric symmetries, generic compact hyperbolic factors have no
continuous isometries, while mixing with a Cartan of a preserved round
$S^2$ would break its non-Abelian $SU(2)$ flavour symmetry.  A torus
translation is subtler because background flux can obstruct its massless
lift.  If $k$ generates a translation on $C_j=T^2$, then
$\lambda_k=\iota_k\omega_j$ is non-trivial in $H^1(T^2)$ and
\begin{equation}
    [\iota_k\mathfrak g]\supset
    N_j\,u_B[\lambda_k]\in u_BH^1(B_6)\subset H^3(M_8).
    \label{torus-flux-obstruction}
\end{equation}
For $N_j\neq0$, this class cannot be cancelled by the independent
$H^3(B_6)$ component, so $\iota_kG_4$ is not globally exact and the standard
combined metric--form massless vector is flux-obstructed
\cite{Bianchi:2010es}.  Every torus factor in the explicit examples of
section~\ref{applications} carries nonzero $N_j$.

These checks do not classify every boundary condition or exclude an accidental
infrared Abelian current, so the absence of additional R-mixing directions remains
an assumption and any enlargement would require a new extremization. Under the
stated boundary prescription, however, the explicit calculations establish
nondegenerate decoupled blocks for the additional harmonic currents and a flux
obstruction for the torus translations, supporting the trial sector used in the
three examples.

\section{Wrapped-M5 bulk calibration details}
\label{M5-details}

This appendix gives the direct sphere-bundle calculation underlying the
normalization check quoted in section~\ref{M5check}.  In particular, it
shows explicitly that the $N^3$ and $N$ coefficients follow from the cubic
and $C_3\wedge X_8$ characteristic terms independently of the six-dimensional
field-theory polynomial.

\subsection{Angular-form reduction on the four-sphere bundle}
\label{M5-bulk-sphere-details}

Let $Z_8=M_4\times W_4$, where $W_4$ supports the external characteristic
classes, and consider the oriented rank-five real bundle
\begin{equation}
 \cN=\RR\oplus(\cL_A)_{\RR}\oplus(\cL_B)_{\RR},
 \qquad c_1(\cL_A)=n_A,\quad c_1(\cL_B)=n_B,
 \qquad \rho:S(\cN)\longrightarrow Z_8.
 \label{M5-sphere-extension}
\end{equation}
The roots include the independent sources of
\eqref{M5-roots-independent}.  The compact internal space is the
restriction of this $S^4$ bundle to $M_4$.  As in the projective calculation,
$S(\cN)$ is a twelve-dimensional characteristic-class extension, not a
claim about a supersymmetric Euclidean filling.

Let $e_4=e(T_\rho)$ be the Euler class of the oriented vertical tangent
bundle.  We use the cohomology class of the normalized global angular form,
with $\int_{S^4}e_4=2$.  The stable identity
\begin{equation}
 T_\rho\oplus\RR\simeq\rho^*\cN
\end{equation}
and $e(T_\rho)^2=p_2(T_\rho)$ imply
\begin{equation}
 e_4^2=\rho^*p_2(\cN),\qquad
 \rho_*e_4=2,\qquad \rho_*e_4^2=0,\qquad
 \rho_*e_4^3=2p_2(\cN).
 \label{M5-angular-identities}
\end{equation}
Here and below these are identities of characteristic classes, not
necessarily equalities of particular differential-form representatives.
They also follow directly by localization.  For the torus rotating the two
complex planes, the north and south sections have oriented vertical Euler
classes $E_+=n_A n_B$ and $E_-=-n_A n_B$, with
$e_4|_\pm=E_\pm$.  Thus,
\begin{equation}
 \rho_*e_4^m=\frac{E_+^m}{E_+}+\frac{E_-^m}{E_-}
 \quad (m=1,2,3),
\end{equation}
which gives \eqref{M5-angular-identities} before specializing the roots.

For the standard wrapped stack, choose the angular flux class
\begin{equation}
 \widehat{\mathfrak G}_{S^4}=\frac{N}{2}e_4,
 \qquad \int_{S^4}\mathfrak g=N>0.
 \label{M5-angular-flux}
\end{equation}
This specifies the wrapped-M5 flux sector.  No additional independent
pulled-back magnetic four-class is turned on. The stable tangent splitting
on the total space gives
\begin{equation}
 T S(\cN)\oplus\RR\simeq\rho^*(TZ_8\oplus\cN),\qquad
 \widehat X_8=\rho^*X_8(TZ_8\oplus\cN).
\end{equation}
The two bulk contributions therefore reduce to
\begin{equation}
 \rho_*\left(\frac16\widehat{\mathfrak G}_{S^4}^{\,3}\right)
 =\frac{N^3}{24}p_2(\cN),\qquad
 \rho_*\left(\widehat{\mathfrak G}_{S^4}\widehat X_8\right)
 =N X_8(TZ_8\oplus\cN).
 \label{M5-bulk-eight-form-details}
\end{equation}
Both coefficients follow from the flux normalization and sphere integration.

We quote the boundary anomaly, with the same characteristic combination
$\frac16\mathfrak G^3+\mathfrak G X_8$ as in
\eqref{lift-local-anomaly}.  In the conventional wrapped-M5 orientation,
$N$ is positive and the supersymmetric current is right-moving. The
boundary polynomial has the opposite sign to the brane inflow polynomial
of \cite{Bah:2019inflow}.  Reversing the two-dimensional orientation
reverses the signed anomaly polynomial and interchanges the left/right
labels.  The relative sign between the cubic and $X_8$ terms is unchanged.
The roots $r\pm f$ assign unit R-charge and zero flavour charge to the
preserved supercharge with normal-plane spin weights $(1/2,1/2)$, fixing
the source normalization independently of the anomaly comparison.

For an explicit coefficient check, put $b=p_1(TM_4)$ and $p=p_1(TW_4)$.
The bulk four-form is
\begin{equation}
\begin{aligned}
 I_{4,\mathrm{bulk}}={}&
 \left[\int_{M_4}\frac{N^3}{24}n_A^2n_B^2\right]_4\\
 &+\frac{N}{192}\left[\int_{M_4}
 \left\{(b+p-n_A^2-n_B^2)^2-4bp-4n_A^2n_B^2\right\}\right]_4.
\end{aligned}
 \label{M5-bulk-root-polynomial}
\end{equation}
Only internal degree four and external degree four are retained.
Pure internal or external degree-eight tangent terms do not contribute.
Substitution of \eqref{M5-roots-independent} gives the coefficients
listed below, with $C=P_1+2\chi$.
\begin{center}
\renewcommand{\arraystretch}{1.5}
\begin{tabular}{c cc}
\toprule
Monomial & Coefficient of $N^3$ & Coefficient of $N$\\
\midrule
$r^2$ & $C(3-z^2)/48$ & $(Cz^2-P_1)/48$\\
$rf$ & $-Cz/12$ & $Cz/12$\\
$f^2$ & $C(3z^2-1)/48$ & $\chi/24$\\
$p$ & $0$ & $-(Cz^2+3P_1+2\chi)/192$\\
\bottomrule
\end{tabular}
\end{center}
The entries are coefficients of monomials in $I_4$, rather than entries of
$k_{AB}$.  A diagonal gauge coefficient is $k_{AA}/2$, whereas the mixed
coefficient is $k_{RF}$. The independently known six-dimensional anomaly is
compared with these coefficients in section~\ref{M5check}. The order-one centre-of-mass subtraction
lies outside the two bulk terms in \eqref{M5-bulk-eight-form-details}, which
nonetheless reproduce separately all the displayed $N^3$ and $N$ gauge and
gravitational coefficients.

\section{Endpoint benchmark for the \texorpdfstring{$(+,+,0)$}{(+,+,0)} background}
\label{projective-benchmark}

We record the radial benchmark also described in appendix~B of
\cite{GregoryNotmanSparks:2026localizing} and evaluate its leading central
charge using endpoint data.  This provides a numerical comparison with the
flux formula, with the scope stated in table~\ref{comparison-scope}.
The radial coordinate in this appendix is $\varrho$, and primes denote its
derivatives. The functions $Q_i$, $P$, $R$ and $D$ below are combinations of
the radial profile $H$, curvature-sign parameters $k_i$ and integration
constants $c_i,d_i$. In the conventions of \cite{Gregory:nonpoly}, for
cyclic $(i,j,k)$, define
\begin{align}
 Q_i&=k_jk_k\varrho^2-4c_i\varrho-4d_i,\qquad P=Q_1Q_2Q_3,\\
 R&=4\sum_i k_i d_i+4\varrho\sum_i k_i c_i
       -3k_1k_2k_3\varrho^2,\qquad D=2H''+R.
\end{align}
The radial equation and regularity conditions are
\begin{equation}
 -(H')^2+HD+\frac14P=0,\qquad
 H(\varrho_\pm)=0,\qquad H>0,\quad Q_i>0,\quad D<0
 \quad (\varrho_-<\varrho<\varrho_+).
\end{equation}
The pole slopes are $H'_- =\sqrt{P_-}/2$ and
$H'_+=-\sqrt{P_+}/2$.  Differentiating the radial equation also imposes
$H'_\pm R_\pm+P'_\pm/4=0$ at each smooth pole.  The printed benchmark is
\begin{align}
 (k_1,k_2,k_3)&=(1,1,0),& (c_1,c_2,c_3)&=(1.02,1.02,0),\\
 (d_1,d_2,d_3)&=(-1,-0.98489221891510,0.13201704744117),\\
 (\varrho_-,\varrho_+)&=(0.76580730923010,0.85624260578730),\\
 (H''_-,H''_+)&\simeq(-2.13402573755,-2.50355915586).
\end{align}
The numerical construction and its regularity diagnostics are described in section~5.3
and appendix~B.1 of \cite{GregoryNotmanSparks:2026localizing}. Here we use only the
published endpoint values, so the calculation checks the endpoint formulae without
repeating the boundary-value integration.

For an explicit endpoint computation, set
\begin{align}
 U_i&=\frac12\left[\frac{H'}{Q_i}\right]_-^+
          -\frac{k_i}{4}(\varrho_+-\varrho_-),\\
 G_i^-&=\frac{H'_-}{2}
       \left(\frac{k_j}{Q_k^-}+\frac{k_k}{Q_j^-}\right)
       -\frac{k_jk_k}{2}\varrho_-+c_i.
\end{align}
With $J_i$ the K\"ahler forms on $C_i$, let
$\nu_i=(4\pi)^{-1}\int_{C_i}J_i$ denote their normalized area
parameters. The overall length scale $\Lambda$, denoted $L$ in the companion
paper, determines the flux normalization
$A=\Lambda^3/(\pi\ell_p^3)$. The magnetic periods are then
\begin{equation}
 N_i=A\nu_i U_i,\qquad M_i=2A\nu_j\nu_kG_i^-.
\end{equation}
Choose $\nu_1=\nu_2=1$, $A=-10/U_1$ and
$\nu_3=-3/(AU_3)$.  The resulting scale and torus area parameter are
\begin{equation}
 A\simeq53.4343928417371,\qquad \nu_3\simeq0.0494059601908484.
\end{equation}
The remaining periods then reproduce $N_2=-11$ and
$(M_1,M_2,M_3)=(10,10,-28)$ to within $3\times10^{-13}$ in absolute value
using the printed data.  This checks the association of the radial
benchmark with the cycle convention in the main text.

The two-derivative Brown--Henneaux integral in these conventions is
\begin{align}
 c_{\rm end}^{(0)}&=\frac38A^3\nu_1\nu_2\nu_3\,\mathcal V,\\
 \mathcal V=-\int_{\varrho_-}^{\varrho_+}D\,\dd\varrho
 &=-2(H'_+-H'_-)-4\sum_i k_i d_i(\varrho_+-\varrho_-)\notag\\
 &\quad-2\sum_i k_i c_i(\varrho_+^2-\varrho_-^2)
       +k_1k_2k_3(\varrho_+^3-\varrho_-^3).
\end{align}
Evaluating these endpoint expressions with the printed decimal inputs gives
\begin{equation}
 \mathcal V\simeq0.5389199886212874,\qquad
 c_{\rm end}^{(0)}\simeq1523.34243709683,
\end{equation}
and hence
\begin{equation}
 \frac{|c_{\rm end}^{(0)}-26215200/17209|}{26215200/17209}
 \simeq2.5\times10^{-14}.
\end{equation}
Section~6.3 of \cite{Gregory:nonpoly} proves that the endpoint and flux expressions
agree algebraically on the endpoint variety. The comparison here checks their numerical
evaluation using the published rounded data, rather than providing bounds on the
interior profile or the supergravity residuals studied in the companion paper. In
particular, the higher-derivative values $(\delta c_L,\delta c_R)=(18,12)$ come from the
retained anomaly polynomial and are not tested by this leading-order endpoint benchmark.

The companion paper gives a stronger, endpoint-based exclusion of
\emph{every} polynomial filling of this benchmark. In the $(+,+,0)$
sector a polynomial solution has degree at most three, and direct
substitution of the unique endpoint-matching cubic yields a nonzero
$t^2(1-t)^2$ residual for every positive torus volume
\cite[equations~(5.3)--(5.4)]{Gregory:nonpoly}. As an additional
flux-space diagnostic, the necessary condition for its regular generic
cubic branch has residual
\begin{align}
 \mathcal R_{\rm poly}
 &=N_1^2M_1(M_1+N_3)(M_1+2N_3)
   -N_2^2M_2(M_2+N_3)(M_2+2N_3)\notag\\
 &\quad+2N_1N_2(M_2+N_3)(M_1+N_3)(M_1-M_2)=-5880.
\end{align}
The nonzero flux residual is a complementary obstruction on that generic
cubic branch. The complete polynomial exclusion is the endpoint-substitution
argument cited above, which is independent of this branch elimination and of
the small numerical deviation from an endpoint-matching polynomial.
The regularity and existence analysis remains that of
\cite{Gregory:nonpoly}.

\section{Conditional Euclidean relative interpretation}
\label{lorentzian-dictionary}

Motivated by the established relative-equivariant M-theory Chern--Simons
formalism \cite{BenettiGenolini:2026odd,BenettiGenolini:2026pathintegral} and related
higher-derivative constructions
\cite{BenettiGenolini:2026higher,MartelliZaffaroni:2024,Gaar:2026index,Cassani:2026toric,BenettiGenolini:2026airy},
we describe an optional relative interpretation of the compact anomaly
calculation in the purely magnetic subclass $f_{\rm el}=0$, with the
choice $a_0=0$. The electric backgrounds admitted in the main text require
an extension of this continuation and disc construction, including the
contraction of the electric flux with the external circle. Such an
extension is not assumed here.
 The
local continuation and disc identities provide the required formal ingredients, but do
not construct the global filling needed for an action interpretation. Such an
interpretation would additionally require a globally free external circle, a spin
filling, smooth extended bilinears and the relative boundary representative specified in
section~\ref{global-euclidean-scope}. These hypotheses play no role in the compact cubic
or $X_8$ push-forward, and the internal supersymmetry lift is recorded independently in
appendix~\ref{internal-lift-details}.

\subsection{External bilinears and continuation}

For completeness we record the normalization that relates the physical
Lorentzian solution to the Euclidean bilinears used in
appendix~\ref{spinor-bridge}. Let $\eta_L$ be an eleven-dimensional
Lorentzian Killing spinor and $\psi_\pm$ its external spinor components.
With a $3+8$ Clifford splitting, write
\begin{equation}
    e^{-\lambda/2}\eta_L=\psi_-\otimes\epsilon+\psi_+\otimes\epsilon^c,
    \qquad
    V_{L\mu}=2\psi_-^TC_3\gamma_\mu\psi_+.
\end{equation}
Here $\psi_\pm=(\psi_1\pm i\psi_2)/\sqrt2$ are formed from real
three-dimensional spinors.  The external gamma matrices obey
$\{\gamma_\mu,\gamma_\nu\}=2\eta_{\mu\nu}$ with
$\eta=\mathrm{diag}(-,+,+)$, and $C_3$ is the antisymmetric charge-conjugation
matrix.  One explicit real choice is
\begin{equation}
 \gamma_0=\begin{pmatrix}0&1\\-1&0\end{pmatrix},\qquad
 \gamma_1=\begin{pmatrix}0&1\\1&0\end{pmatrix},\qquad
 \gamma_2=\begin{pmatrix}-1&0\\0&1\end{pmatrix},\qquad C_3=\gamma_0.
 \label{external-clifford-representation}
\end{equation}
With $\Gamma_\mu=\gamma_\mu\otimes\gamma_9$ and
$\Gamma_m=-\mathbf1\otimes\gamma_m$, the Clifford orientation is
$\Gamma_{0\cdots10}=-1$, as in
\cite[appendix~A]{Ashmore:2023}.  We take
$C_{11}=C_3\otimes\gamma_9$ and define the eleven-dimensional forms by
$K_L^\flat=\eta_L^TC_{11}\Gamma_{(1)}\eta_L$ and
$\Omega_L=\eta_L^TC_{11}\Gamma_{(2)}\eta_L$, using the physical warped
coframe in $\Gamma_{(p)}$.  The minus sign in $\Gamma_m$ fixes the internal
one-form and mixed two-form signs below.  Define the external two-form by
\begin{equation}
 Q_L=2\psi_-^TC_3\gamma_{(2)}\psi_+,\qquad
 \gamma_{(2)}=\tfrac12\gamma_{\mu\nu}e^\mu\wedge e^\nu.
 \label{external-Q-definition}
\end{equation}
It is dual to $V_L$, with the sign fixed by this Clifford orientation.
In particular, $\iota_{V_L^\sharp}Q_L=0$.  $Q_E$ denotes its analytic
continuation with the same external frame prescription used for $V_L$.
No normalization of $Q_E$ enters the internal anomaly push-forward.

If $a=\psi_1^TC_3\psi_2$, the two-component Fierz identity gives
$V_L^2=-4a^2$, while the eleven-dimensional bilinears reduce to
\begin{equation}
    e^{-2\lambda}K_L^\flat=V_L-2aL,
    \qquad
    e^{-3\lambda}\Omega_L=\zeta Q_L-V_L\wedge\kappa+2a\omega.
\end{equation}
Demanding that the internal projection of $K_L$ equal the canonical
R-symmetry vector $\xi^\flat=L/2$ fixes $a=-1/4$.  Hence
\begin{equation}
    V_L^2=-\frac14,
    \qquad
    K_L=V_L^\sharp+\xi,
    \qquad
    \Omega_L=e^{3\lambda}
    \left(\zeta Q_L-V_L\wedge\kappa-\frac12\omega\right).
\end{equation}
Wick rotating only the external timelike bilinear,
\begin{equation}
    V_L\longrightarrow V_E=i\,v,
    \qquad v^2=\frac14,
\end{equation}
with all internal forms fixed gives
\begin{equation}
    K_E=i\,v^\sharp+\xi,
    \qquad
    \Omega_E=e^{3\lambda}
    \left(\zeta Q_E-i\,v\wedge\kappa-\frac12\omega\right),
    \label{KE-final}
\end{equation}
which are used in appendix~\ref{spinor-bridge}.  Since the Euclidean spinor
is complex, the imaginary external component is compatible with Euclidean
supersymmetry.  No Lorentzian reality condition is imposed after the
continuation.

\subsection{Euclidean relative-equivariant formulation}
\label{relative}

We specialize \eqref{lorentzian-input} to $f_{\rm el}=0$ and choose
$a_0=0$, so that $G_L=e^{3\lambda}F$ and
$\mu=\tfrac12e^{3\lambda}\zeta$ in this appendix.
  We keep the internal metric, flux and warp
factor fixed under continuation.  The relative-equivariant interpretation of the
Chern--Simons sector is instead organized in Euclidean eleven-dimensional
supergravity, following \cite{BenettiGenolini:2026odd}.  The Euclidean
spinor is complex and the supersymmetry bilinears obey
\begin{equation}
    \dd_{K_E}\Phi_E^G=0,
    \qquad
    \Phi_E^G=G_E+\Omega_E,
    \qquad
    \dd_{K_E}=\dd-\iota_{K_E}.
    \label{euclidean-PhiG}
\end{equation}
The symbols $\Phi_{W,E}^{G}$ and $\Phi_{W,E}^{X_8}$ below denote
equivariant extensions of the flux polyform and the characteristic form to
the assumed Euclidean filling, respectively. The anomaly form through the
protected eight-derivative Chern--Simons term is
\begin{equation}
    \Phi_E^{(\mathrm{anom})}
    =\frac{i}{6}\left(\Phi_{W,E}^{G}\right)^3
    +i(2\pi\ellp)^6\Phi_{W,E}^{G}\wedge\Phi_{W,E}^{X_8}.
    \label{anom-polyform-audited}
\end{equation}
The first term enters the relative localization of the two-derivative on-shell
action. The second is used only through equivariant integration of the known
characteristic class over the closed compact $M_8$, without assuming a complete
higher-derivative localization theorem. This keeps the higher-derivative anomaly
calculation anchored in the compact characteristic-class push-forward.

\subsubsection{Analytic continuation and the internal polyform}
\label{spinor-bridge}

The continued bilinears are given in \eqref{KE-final}, with
$\iota_{v^\sharp}Q_E=0$.  The internal magnetic polyform
\eqref{AdS3-flux-polyform} and its closure follow from the identities in
appendix~\ref{internal-lift-details}.  They fix the internal data used in
the local disc construction below.

\subsubsection{Disc push-down and fixed sections}
\label{disc-bridge}

Let the real free external circle be generated by
$\ell_{\rm ext}=qv^\sharp$, with $q$ fixed so that this generator
has period $2\pi$. Write its angular coordinate as
$\varphi\sim\varphi+2\pi$, and choose the real connection one-form
$\alpha_{\rm ext}$ with $\iota_{\ell_{\rm ext}}\alpha_{\rm ext}=1$.
Thus, $\int_{S^1}\alpha_{\rm ext}=2\pi$, and its curvature has
integral periods $\frac1{2\pi}\int_{\mathcal C}\dd\alpha_{\rm ext}
\in\ZZ$ over any closed two-cycle $\mathcal C$ of the circle base.
These conventions identify $\dd\alpha_{\rm ext}/(2\pi)$ with the
first Chern form of the circle bundle. Since the external component of
$K_E$ is $iv^\sharp$,
\begin{equation}
    w\equiv\iota_{K_E}\alpha_{\rm ext}=\frac{i}{q}.
    \label{euclidean-w}
\end{equation}
For a $\dd_{K_E}$-closed polyform $\Psi$ satisfying
$\iota_{\ell_{\rm ext}}\Psi=-\dd_{K_E}\sigma$, let $\rho\in[0,1]$
be a radial coordinate on the filling disc, with $\rho=0$ at its centre.
The canonical push-down and disc lift are
\begin{equation}
\begin{aligned}
    \Psi_B&=\Psi-\dd_{K_E}(\alpha_{\rm ext}\wedge\sigma),\\
    \Psi_W&=\Psi_B+\dd_{K_E}(\rho^2\alpha_{\rm ext}\wedge\sigma).
\end{aligned}
\label{disc-pushdown}
\end{equation}
Using $v^2=1/4$, $\iota_{v^\sharp}Q_E=0$ and
$\dd\mu=e^{3\lambda}\kappa$, contraction of
\eqref{KE-final} gives
$\iota_{qv^\sharp}\Omega_E=-iqe^{3\lambda}\kappa/4$.
Hence
\begin{equation}
    \iota_{\ell_{\rm ext}}\Phi_E^G
    =-\frac{iq}{4}\dd\mu,
    \qquad
    \sigma_E=\frac{iq}{4}\mu,
    \qquad
    w\sigma_E=-\frac{\mu}{4}.
    \label{euclidean-wsigma}
\end{equation}
Hence the internal part of the Euclidean push-down is exactly
\begin{equation}
    \left.\Phi_{B,E}^{G}\right|_{M_8}
    =e^{3\lambda}F-\frac12e^{3\lambda}\omega-\frac{\mu}{4}
    =\Phi^F.
    \label{PhiBE-equals-PhiF}
\end{equation}
At the disc centre the external circle collapses, leaving as fixed components precisely
the fixed sets of $\xi$ on $M_8$. For the projective geometries under consideration,
these are the two pole sections, denoted $\mathcal F_\pm$. Write
$N\mathcal F_\pm$ for their full normal bundles and $L(D)$ for the
complex normal line to the collapsed external circle. Their equivariant
Euler classes are
\begin{equation}
    N\mathcal F_\pm=L(D)\oplus N_\pm,
    \qquad
    e_{K_E}(N\mathcal F_\pm)
    =e_D\,e_\xi(N_\pm),
    \qquad
    e_D=\frac{w-\dd\alpha_{\rm ext}}{2\pi}
       =-\frac{\dd_{K_E}\alpha_{\rm ext}}{2\pi}.
    \label{euler-projective-relative}
\end{equation}
The local Euclidean relative construction therefore reproduces both the internal flux
polyform and the projective Euler factor that enter the compact calculation in the main
text.

\subsubsection{Relative interpretation of the cubic push-forward}
\label{relative-cubic-disc}

Assume the global disc construction exists and the relative localization
hypotheses hold.  In the two-derivative theory the relevant pair consists
of the anomaly polyform $\Phi_E^{(\mathrm{anom},0)}$ and a gauge-invariant
boundary representative $\Phi_E^{(\mathrm{GI})}$ satisfying
\begin{equation}
 \dd_{K_E}\Phi_E^{(\mathrm{GI})}
 =-\left.\Phi_E^{(\mathrm{anom},0)}\right|_{M_{11}^E}.
 \label{relative-boundary-representative}
\end{equation}
Relative localization applies to their combined functional
\cite{BenettiGenolini:2026odd},
\begin{equation}
 I_E^{\mathrm{rel},(0)}
 =\frac1{16\pi G_{11}}\left[
 \int_{M_{11}^E}\Phi_E^{(\mathrm{GI})}
 +\int_{W_{12}}\Phi_E^{(\mathrm{anom},0)}\right],
 \label{relative-functional}
\end{equation}
where each integral selects the top-degree component.  Let $\Sigma_2$ be
the base of the external circle.  With fixed components
$\mathcal F_\pm\simeq\Sigma_2\times B_6^\pm$, the factorization of their
Euler classes gives
\begin{equation}
\begin{aligned}
 I_E^{\mathrm{rel},(0)}
 &=2\pi i\sum_\pm\int_{\Sigma_2\times B_6}
   \frac{\widehat{\mathfrak G}_\pm^3/6}{e_D\,e_\xi(N_\pm)}\\
 &=2\pi i\int_{\Sigma_2}\frac{\mathcal P_4}{e_D},
 \qquad e_D=-\frac{\dd_{K_E}\alpha_{\rm ext}}{2\pi}.
 \label{full-relative-cubic}
\end{aligned}
\end{equation}
Here $\mathcal P_4$ is the compact push-forward defined in
section~\ref{cubic-relative-fixed-point}, evaluated on the
background sources of the assumed filling. The identical normalization
of $e_D$ in \eqref{euler-projective-relative} and
\eqref{full-relative-cubic} follows from the $2\pi$-periodic circle
connection, with curvature divided by $2\pi$ rather than left
unnormalised.  The relative boundary term in
\eqref{relative-functional} is essential.  The bare Chern--Simons
integral does not generally equal this fixed-point sum on a disc bundle
with boundary.  The compact anomaly polynomial in the main text is
computed independently of this relative interpretation.

\subsubsection{Global scope}
\label{global-euclidean-scope}

The preceding identities are local and do not themselves supply the global filling.
Realizing the disc construction requires a Euclidean $M_{11}^E$, a spin filling
$W_{12}$, a globally free external circle, smooth extensions of the bilinears and the
appropriate relative boundary representative. For the projective topology, the natural
candidate is
\begin{equation}
    P\!\left(\cO\oplus\cL\boxtimes\cR^2\right)
    \longrightarrow B_6\times W_4,
\end{equation}
and its spin condition is verified in appendix~\ref{projective-technical}. We have not
constructed a global supersymmetric representative of this model for every Lorentzian
background, so the relative interpretation remains conditional on such a realization.
The cubic and $X_8$ polynomials used throughout the main text, by contrast, follow
directly from equivariant integration over the closed $M_8$.





\end{document}